\documentclass[12pt]{article}

\usepackage{lscape,bbold}

\usepackage[final]{changes} 
\usepackage{setspace,graphicx,epstopdf,amsmath,amsfonts,amssymb,amsthm,versionPO}
\usepackage{marginnote,datetime,enumitem,subfigure,rotating,fancyvrb}
\usepackage{hyperref,float}
\usepackage[longnamesfirst]{natbib}
\usepackage{multirow,bigdelim}
\usepackage{graphicx,subfigure}
\usdate

\excludeversion{notes}		
\includeversion{links}          

\iflinks{}{\hypersetup{draft=true}}

\usepackage[margin=1in]{geometry}%
\usepackage{todonotes}%

\makeatletter\let\chapter\@undefined\makeatother 

\usepackage{adjustbox}
\usepackage{array}
\newcolumntype{R}[2]{%
    >{\adjustbox{angle=#1,lap=\width-(#2)}\bgroup}%
    l%
    <{\egroup}%
}
\newcommand*\rot{\multicolumn{1}{R{60}{1em}}}

\usepackage{indentfirst} 
\usepackage{jfe}          

\begin{document} 

\title{\vspace*{-1cm}{\Large Modelling Sovereign Credit Ratings: Evaluating the Accuracy and Driving Factors using Machine Learning Techniques\thanks{%
The authors would like to thank Dick van Dijk, Hüseyin Özturk and Corne Vriends for their comments and suggestions, as well as three anonymous referees.}
}}
\author{Bart H.L. Overes\thanks{%
Corresponding author. Mail: \texttt{Overes.Bart@gmail.com}.} \\
Erasmus University Rotterdam \and Michel van der Wel \\
Erasmus University Rotterdam, ERIM, Tinbergen Institute, and CREATES}

\renewcommand{\thefootnote}{\fnsymbol{footnote}}

\singlespacing

\maketitle

\begin{abstract}
Sovereign credit ratings summarize the creditworthiness of countries. These ratings have a large influence on the economy and the yields at which governments can issue new debt. This paper investigates the use of a Multilayer Perceptron (MLP), Classification and Regression Trees (CART), \added{Support Vector Machines (SVM), Naïve Bayes (NB),} and an Ordered Logit (OL) model for the prediction of sovereign credit ratings. We show that MLP is best suited for predicting sovereign credit ratings, with a random cross-validated accuracy of 68\%, followed by CART (59\%), \added{SVM (41\%), NB (38\%),} and OL (33\%). Investigation of the determining factors shows that \added{there is some heterogeneity in the important variables across the models.}\deleted{roughly the same explanatory variables are important in all models,} \added{However, the two models with the highest out-of-sample predictive accuracy, MLP and CART, show a lot of similarities in the influential variables,} with regulatory quality, \added{and} GDP per capita\deleted{, and unemployment rate} as common important variables. Consistent with economic theory, a higher regulatory quality and/or GDP per capita are associated with a higher credit rating\deleted{, while a higher unemployment rate is associated with a lower credit rating}.


\end{abstract}

\thispagestyle{empty}

\clearpage

\newpage
\onehalfspacing
\setcounter{footnote}{0}
\renewcommand{\thefootnote}{\arabic{footnote}}
\setcounter{page}{1}

\section{Introduction} 
\label{sec:intro}
A sovereign credit rating is an evaluation of the credit risk of a country and gives an indication of the likelihood that the country will be able to make promised payments. These ratings have a large influence on the interest rate at which governments are able to issue new debt and thereby a big effect on government spending and the government deficit. Sovereign credit ratings are usually given by one of three Credit Rating Agencies (CRAs): Moody's, S\&P, and Fitch. These agencies use a combination of objective and subjective factors to determine the rating, however, unfortunately, the exact rating methodology and the determining factors remain unknown. This lack of transparency has resulted in widespread criticism of the CRAs. They have, among other things, been accused of giving biased ratings \citep{Luiten_2016}, reacting slowly to changing circumstances \citep{Elkhoury_2009}, and behaving procyclically \citep{Ferri_1999}.\\
\indent Getting an understanding of the rating methodology and the determining factors would be very helpful for governments, investors, and financial institutions. Governments would be able to anticipate possible rating changes, while investors and financial institutions could check if ratings deviate from what the fundamentals of a country imply. In order to get an understanding of the credit rating process, a model is needed that can predict the ratings, ideally with high accuracy. Research has, up until now, mostly focussed on modelling sovereign credit ratings using various forms of the Ordered Probit/Logit (OP/OL) model, which assumes a particular functional form for the relation between a linear combination of the input variables and the continuous output variable, or other related models, see, for example, \cite{Cantor_1996,Dimitrakopoulos_2016,Reusens_2017}.\footnote{For ease of reference, we will refer to the probit/logit variants simply as linear forms because of the linear relation among variables.} These models allow for easy interpretation of the determining factors and prove to be fairly accurate, but come at the cost that the linear relation they assume might not always hold. A recent branch of research has therefore focussed on using Machine Learning (ML) techniques to model sovereign credit ratings \citep{Bennell_2006, Ozturk_2015,Ozturk_2016}. \cite{Ozturk_2015,Ozturk_2016} show that ML models outperform linear models on predictive accuracy, sometimes by a large margin. \added{Multilayer Perceptron (MLP), Classification and Regression Trees (CART), Support Vector Machines (SVM), and Naïve Bayes (NB) are among the commonly used techniques. Where} especially \deleted{the Multilayer Perceptron (}MLP\deleted{)} and \deleted{Classification and Regression Trees (}CART\deleted{)} prove to be well suited for modelling sovereign credit ratings. However, getting an insight into the inner workings of the models and their determining factors is difficult.\\
\indent This paper focusses on obtaining the determining factors of \deleted{two}\added{four} ML models used for sovereign credit rating\deleted{s}; MLP\replaced{,}{and} CART, \added{SVM, and NB, the latter of} which has, up until now, not been done for ML models in the sovereign credit rating setting. This will give insight into the way in which the ML models give the ratings and what variables are important in the process, lack of these insights has been the main weakness of the ML models to date. In order to obtain the determining factors, we use so called Shapley Additive exPlanations (SHAP) \citep{Lundberg_2017}. SHAP allow for the isolation of each variable's effect and can pick up on non-linear relations, making them well suited for the interpretation of ML models. Getting an understanding of these more accurate models will help figure out the driving factors and methodologies for sovereign credit ratings, as interpreting a model is only useful when that model accurately represents reality. We contrast these approaches to an OL model, this allows for examining how different the insights are for Machine Learning methods compared to a more econometric approach. \\
\indent This study uses Moody's credit ratings for a set of 62 developed and developing countries, such as Argentina, China, Germany and New Zealand, for the period 2001-2019, to train and evaluate the models. The explanatory variables are similar to those of \cite{Dimitrakopoulos_2016}: GDP growth, inflation, unemployment, current account balance, government balance, government debt, political stability, regulatory quality, and GDP per capita. \added{These variables are chosen because they proved to be important in the credit rating process in earlier studies \citep{Cantor_1996,Afonso_2003, Butler_2006}.}\\
\indent We document that MLP is the most accurate model for sovereign credit ratings with an accurate rating prediction \added{in a random split cross-validation} of 68\%, and 86\% of ratings correct within 1 notch. Where the percentage within 1 notch indicates what fraction of the ratings given by the model does not deviate more than 1 class from the actual rating. CART follows relatively closely with an accuracy of 59\%, and 76\% correct within 1 notch. \added{The other two Machine Learning techniques, SVM, with 41\% correct and 59\% within 1 notch, and NB, 38\% correct and 61\% within 1 notch, prove to be less accurate.} OL \deleted{, however,}significantly underperforms \added{the best ML techniques,} with correct predictions for only 33\% of the observations, and 57\% within 1 notch. Analysis of the determining factors shows \added{some heterogeneity between the different modelling techniques. Nonetheless,} \deleted{that}regulatory quality and GDP per capita are very important explanatory variables in \added{the two best performing models, being MLP and CART.}\deleted{the credit rating process, especially for the MLP and CART. The unemployment rate also proves to be influential in all the models.} The relation between these explanatory variables and the credit rating is as expected\deleted{ in all models}, with regulatory quality and GDP per capita having a positive influence \added{on the credit rating}\deleted{and unemployment rate having a negative influence}.\\
\indent The structure of our paper is as follows. We begin by discussing the methodology used in this study in Section \ref{sec:meth}, directly followed by a discussion of the data in Section \ref{sec:data}. Section \ref{sec:results} gives an overview of the results obtained in this study. Section \ref{sec:concl} concludes.

\section{Methodology}
\label{sec:meth}

\noindent In this section, we discuss the methods used in this study, starting off with the modelling techniques. Thereafter, we discuss the so called SHAP values, which allow us to isolate the influence of individual variables in complex models. Lastly, the methods used to evaluate and compare the accuracy of the different models are discussed.

\subsection{Modelling techniques}
\label{sec:modellingtechniques}

This section gives an overview of the different models. These models are used to predict the sovereign credit rating denoted by $y_i$ for observation $i$, which represents a rating class, with $m$ being the total number of classes. In this research we use Moody's credit ratings, Moody's gives categorical credit ratings ranging from Aaa (highest) to C (lowest), with 19 categories in between. As algorithms in general cannot handle categorical ratings, they are transformed to numeric ratings from 17 (Aaa) to 1 (Caa1 and lower), where all the ratings of Caa1 and lower have been grouped in $C_{combined}$ because of their infrequent occurrence. The structure of $y_i$ therefore is as follows:
\begin{equation}
y_{i} = \begin{cases}
Aaa \ (17)  \\ 
Aa1 \ (16)  \\
\quad \vdots\\
C_{combined} \ (1) 
\end{cases}
\end{equation}
\noindent with the numerical value corresponding to a rating given in brackets. Thus, a high numerical value corresponds to a high credit rating. The explanatory variables are contained in $X_i$, and $n$ is the total number of observations. Consistent with the main modelling approaches in the literature that we follow (see, e.g, \cite{Dimitrakopoulos_2016,Ozturk_2015,Ozturk_2016}), the panel structure of the credit rating data is not taken into account. The main reason for not using a panel structure is that the ML models used in this study do not support panel data, although there are developments in this area.

\paragraph{Multilayer Perceptron} \mbox{} \\
The Multilayer Perceptron (MLP) is a form of an Artificial Neural Network (ANN) which mimics the way that the human brain processes information. MLPs, or similar Neural Network type of algorithms, are often found to perform very well in classification problems involving corporate and sovereign credit rating, see, for example, \cite{Baessens_2003,Lessman_2015,Ozturk_2015,Ozturk_2016}. A MLP is able to model non-linearities in the data, and can therefore handle very complex classification problems with heterogeneous groups. However, interpretation of the MLP is extremely difficult and, up to a certain degree, it will always remain a ``black box''. \\
\indent The MLP consists of an input layer, an output layer and a certain number of hidden layers in between. The input layer contains a number of neurons equal to the number of explanatory variables, here the set of explanatory variables ($X_i$) are fed into the model. This layer is followed by a certain number of hidden layers, which contain neurons that get an input signal from all the neurons in the previous layer and process that information in order to generate an output that is passed on to every neuron in the next layer. The output layer is the final layer in the MLP structure and has a number of neurons equal to the number of desired outputs, in this case the probability of belonging to each of the credit rating categories in $y_i$.\\
\indent The output for each neuron $j$ in hidden layer $i$ is given by
\begin{equation}
\label{eq:MLPneuronoutput}
h^{(i)}_{j} = \sigma^{(i)}\left(z^{(i)}_j\right) = \sigma^{(i)}\left(b^{(i)}_{j} + \sum_{k = 1}^{n^{(i-1)}} W^{(i-1)}_{jk} \cdot h^{(i-1)}_{k}\right),
\end{equation}
\noindent where $b^{(i)}_{j}$ presents the bias term, $W^{(i-1)}_{jk}$ gives the weight connecting neuron $k$ from layer $i-1$ to neuron $j$ in layer $i$ and $n^{(i-1)}$ is the total number of neurons in layer $i-1$. The activation function $\sigma^{(i)}(z)$ enables the algorithm to model non-linearities that might be present in the data, and can be varied for each layer \citep{Baessens_2003}. We use a Rectified Linear Unit (ReLU) function for the hidden layers, given by
\begin{equation}
\label{eq:ReLU}
\sigma(z) = max(0,z),
\end{equation}
since it is often found to perform best \citep{Rmachandran_2017}. For the output layer, we use the Softmax function, given by
\begin{equation}
\label{eq:Softmax}
\sigma(z_j) = \frac{e^{z_j}}{\sum_{k=1}^{m}e^{z_k}} \quad \textrm{for} \ j = 1,...,m \ \textrm{and} \ z = (z_1,...,z_m) ,
\end{equation}
where $m$ is the number of desired output categories. This activation function gives us, for every country, the probability of belonging to each rating class and is therefore well suited for multiclass classification problems. In the end, the estimate for every country $\hat{y}_{i}$ is set to the numerical class for which it has the highest probability.\\
\indent The MLP is optimized by minimizing the categorical cross-entropy function, given by
\begin{equation}
\label{eq:categoricalcrossentropy}
C(y_{i,j},\hat{y}_{i,j}) = -\sum_{j=1}^{m}\sum_{i=1}^{n}\bigg(y_{i,j}\cdot ln(\hat{p}_{i,j})\bigg)
\end{equation}
\noindent where the number of categories in $y_i$ is given by $m$ and the total amount of observations by $n$. The true value of observation $i$ for class $j$ is given by $y_{i,j}$, which is 1 if observation $i$ belongs to class $j$ and 0 otherwise. The predicted probability that observation $i$ belongs to class $j$  is given by $\hat{p}_{i,j}$. The algorithm is trained by backward propagation of error information through the network. That is, the partial derivative of the cost function with respect to all weights and biases is determined. Thereafter, the weights connecting all the nodes, and the biases are adjusted in such a way that the cost is minimized.\\
\indent The MLP architecture, that is, the number of hidden layers and neurons per hidden layer is optimized through a grid search. In this grid search, we also determine the optimal dropout rate, which is the fraction of neurons dropped at random to prevent overfitting to the training data. The optimal performance-complexity trade-off for this data set is given by the MLP with 1 hidden layer, 256 neurons and a dropout rate of 0.1. Estimation of this MLP is done using a batch size of 8 and 400 epochs. Details of the grid searches can be found in Appendix \ref{app:MLPoptimization}. The MLP is implemented in Python's Keras package \citep{Keras}.

\paragraph{Classification and Regression Trees}\mbox{} \\
\noindent The idea behind the Classification and Regression Tree (CART) is quite simple, the algorithm finds the optimal splits based on the values of the explanatory variables in order to classify the observations. CARTs have shown to be well suited for credit rating, see, for example \cite{DeMoor_2018,Ozturk_2015,Ozturk_2016}. A few of the advantages of CARTs are that they can handle outliers, do automatic feature selection and allow for easy interpretation of the model. However, CARTs can be very prone to overfitting.\\
\indent A CART consists of a root, one or more nodes and several leaves. The first split of the data, based on one of the explanatory variables in $X_i$, is made at the root, that split leads either to a node, where the remaining data is split further, again based on one of the explanatory variables in $X_i$, or a leaf, meaning a decisions is made for these observations. Every observation moves through the tree until it ends up at a leaf, which in our case represents one of the different rating categories in $y_i$. \\
\indent In this research, we use an algorithm that splits the data in two at every node. The sequential data splits are determined using the Gini method. That is, for each variable the algorithm calculates the weighted average Gini impurity, e.g. how effective the different categories can be separated based on that variable, using the following formula
\begin{equation}
\label{eq:Gini}
Gini =\sum_{j=1}^{2} \Bigg ( \frac{n_j}{n_n} \sum_{i=1}^{m} p(i)*(1-p(i)) \Bigg ),
\end{equation}
\noindent where $m$ is the number of different categories in $y_i$ and $p(i)$ is the probability of picking a data point of class $i$ within that branch of the split. Furthermore, $n_j$ is the number of data points assigned to branch $j$ and $n_n$ gives the total number of data points entering that node. The split that leads to the largest decrease in Gini Impurity is used at that node. This means that the CART is greedy, i.e. it does not care about future splits and does not take them into account.\\
\indent CARTs are notorious for overfitting, and therefore sometimes need to be restricted. There are two ways of doing this: restricted growth and pruning. In the case of restricted growth, constraints that limit the growth of the tree in certain ways are implemented, which prevent it from overfitting. Whereas with pruning, the tree is left to grow unrestricted and is decreased in size afterwards. Both methods show no improvement on the cross-validated out-of-sample accuracy, and thus an unrestricted CART is used in this study. The details of the CART optimization can be found in Appendix \ref{app:CARToptimization}. The CART is implemented in Python's scikit-learn package \citep{scikitlearn}.

\paragraph{Support Vector Machines}\mbox{} \\
Support Vector Machines (SVM) is a relatively new Machine Learning (ML) technique which can be used for classification problems. A SVM tries to find the optimal boundary between classes based on the explanatory variables. That is, it constructs a hyperplane which separates the different classes as much as possible. By using a kernel, to transform the data, the SVM can handle non-linear classification problems and is thus not restricted to linear relations. Accuracy of the SVM can be very sensitive to the data used \citep{Hastie_2009}. \cite{Ozturk_2015} find that SVM performs poorly compared to other ML techniques, while it has also shown to be able to match the other ML methods on performance \citep{Ozturk_2016}. \\
\indent The optimal hyperplane for the SVM in a two-class setting is obtained by the following dual optimization problem
\begin{equation}
\begin{split}
\label{eq:SVM}
\max L = \sum_{i=1}^{n}\alpha_i - \frac{1}{2}\sum_{i=1}^{n}\sum_{k=1}^{n}\alpha_{i}\alpha_{k}d_{i}d_{k}K(x_i,x_k) \\
\textrm{s.t.} \quad 0 \leq \alpha_i \leq C, \quad \sum_{i=1}^n\alpha_{i}d_{i} = 0
\end{split}
\end{equation}
\noindent where $n$ is the number of observations, $d_i \in \{-1,1\}$ gives the class label for observation $i$, $\alpha$ is the set of Lagrange multipliers, $x_i$ contains the explanatory variables for observation $i$, and $K(x_i,x_k)$ is the kernel. This two-class optimization problem can be used in a multi-class setting, by optimizing a one-versus-the-rest problem for every class individually. In that case, we get the following 
\begin{equation}
d_{i} = \begin{cases}
-1 \   & \textrm{ if } y_{i}=j\\ 
1 \  & \textrm{ if } y_{i}\not= j\\

\end{cases}
\end{equation}
where $y_i$ is the class to which observation $i$ belongs and $j$ is one of the $m$ classes (17 in this case). By repeating this procedure for all $m$ classes we obtain hyperplanes to separate multiple classes. For further details on the optimization of a SVM and derivation of the dual optimization problem see \cite{Hastie_2009}. In this study we use the Radial Basis Function (RBF) kernel, given by
\begin{equation}
  K(x_i,x_k) = exp(-\gamma\Vert x_{i}-x_{k} \Vert^2),
\end{equation}
\noindent which is the default and recommended kernel for SVM \citep{Liu_2011}. \\
\indent Optimizing a SVM is mainly done by tuning two hyperparameters $C$ and $\gamma$. Of these, $C$ determines the cost of misclassification, where a high $C$ leads to severe punishment of misclassifications, while a low $C$ allows the model to misclassify when determining the optimal hyperplane. The hyperparameter $\gamma$ determines how far the influence of a single training point reaches. When $\gamma$ is low, similarity regions are large, therefore, more points are grouped together, and vice versa for high $\gamma$ values. Optimal settings found in this study are $C = 100,000$ and $\gamma = 10^{-7}$ which are determined using a grid search. Further details of the SVM grid search can be found in Appendix \ref{app:SVMoptimization}. The SVM is implemented in Python's scikit-learn package \citep{scikitlearn}.

\paragraph{Naïve Bayes}\mbox{} \\
The Naïve Bayes (NB) classifier is a Bayesian type of classifier that (naïvely) assumes feature independence. This assumption is very strong, and might not hold in many cases, however, NB handles large predictor sets well, is computationally fast, and is robust to poor data quality \citep{Kotsiantis_2006}. NB has been applied to sovereign as well as corporate credit rating problems with mixed results, see, for example, \cite{Baessens_2003,Lessman_2015,Ozturk_2015,Ozturk_2016}.\\
\indent The NB classifiers determines the probability of observations $i$ belonging to class $j$ ($p_{i,j}$) given the explanatory variables for observations $i$ ($X_i$) using Bayes rule 
\begin{equation}
\label{eq:Naivebayes}
\begin{split}
    p_{i,j} = P(Y_j \mid X_{i}) &= \frac{P(X_{i} \mid Y_j) \, P(Y_j)}{P(X_{i})} \\ &= \frac{P(X_{i} = (x_{i,1}, x_{i,2}, ..., x_{i,n}) \mid Y_j) \, P(Y_j)}{P(X_{i})} \\ &= \frac{P(Y_j) \,  \prod_{k=1}^{n}P(x_{i,k} \mid Y_j)   }{P(X_{i })},
\end{split}
\end{equation}
\noindent where observations $i$ is assigned to the class for which it has the highest probability. Since all the variables used in this study are continuous, we use a simple Gaussian distribution for all variables. \\
\indent As the NB classifier assumes independence of the features, performance can sometimes suffer when correlations between the explanatory variables are high. We therefore iteratively remove highly correlated variables to see if performance increases. The NB is implemented in Python's scikit-learn package \citep{scikitlearn}.

\paragraph{Ordered Logit}\mbox{} \\
\noindent The Ordered Logit (OL) model is, together with the Ordered Probit model, the most frequently used model in literature \citep{Dimitrakopoulos_2016, Afonso_2011,Reusens_2017}. It therefore provides a good benchmark for the Machine Learning models, because these more complex models are only useful when they are able to outperform the OL model. As opposed to OLS, the OL model can deal with unequal distances between rating classes and the presence of a top and bottom category. Furthermore, the OL model allows for interpretation and significance testing of the explanatory variables' coefficients, which  makes it easy to obtain the determining factors.\\
\indent A pooled OL model is implemented. Here, the latent continuous variable $y_{i}^{*}$ has the following specification
\begin{equation}
\label{eq:OPmodel}
y_{i}^{*} = \alpha + X_{i}'\beta + \epsilon_{i},
\end{equation}

\noindent where the intercept is given by $\alpha$, $X_{i}$ contains the explanatory variables for data point $i$, $\beta$ is a vector containing the coefficients and the idiosyncratic errors are given by $\epsilon_{i}$, which has a standard logistic distribution. \\
\indent However, rating categories are not continuous and our continuous variable therefore needs to be transformed into a categorical rating using

\begin{equation}
y_{i} = \begin{cases}
Aaa \ (17)  & \textrm{ if } y_{i}^{*} \geq \tau_{16}\\ 
Aa1 \ (16)  & \textrm{ if } \tau_{16} > y_{i}^{*} \geq \tau_{15}\\
\quad \vdots\\
C_{combined} \ (1) & \textrm{ if } \tau_{1} > y_{i}^{*}
\end{cases}
\end{equation}
\noindent where the boundaries between the different classes are given by $\tau_j$. The OL model is implemented in Python's Mord package \citep{mord}.

\subsection{SHAP values}
\label{sec:SHAPvalues}

\noindent Getting insight into the inner workings of complex models is difficult. Therefore, \cite{Lundberg_2017} came up with a method to approximate the effects that the individual explanatory variables have on the model outcome, called SHAP. This method, based on Shapley values \citep{Shapley_1953}, evaluates how model outcomes differ from the baseline by tuning all the explanatory variables individually, or in combination with a selection of other explanatory variables, while keeping the others constant. \\
\indent The basic framework for explaining a model $f(x)$ using SHAP values is the explanation model
\begin{equation}
\label{eq:SHAPframework}
g(x) = \phi_0 + \sum_{i=1}^{n_{input}}\phi_ix_i,
\end{equation}
\noindent where $x$ is a vector containing all the explanatory variables, $\phi_0$ is the baseline prediction, $\phi_i$ is the weight of the $i^{th}$ explanatory variable in the final prediction, and $n_{input}$ is the total number of explanatory variables. The explanation model $g(x)$ gives an approximation of the output of the real model $f(x)$ by using a linear combination of the input variables and a baseline prediction. Calculating the contribution of each variable $x_i$ to the explanation model $g(x)$ is done using
\begin{equation}
\label{eq:SHAPhpi}
\phi_i(f(x),x) = \sum_{v \subseteq x}^{}\underbrace{\frac{|v|!(n_{input}-|v|-1)!}{n_{input}!}}_{no. \ of \ permutations}\underbrace{\left(f_x(v)-f_x(v\backslash i) \right)}_{contribution \ of \ i}, 
\end{equation}
\noindent where, $v \subseteq x$ represents all the possible $v$ vectors where the non-zero elements are a combination of the non-zero elements in $x$, $f_x(v\backslash i)$ is the model output of the original model with the $i^{th}$ element of $v$ set to zero, and $|v|$ gives the total number of non-zero elements in $v$ \citep{Lundberg_2017}. The SHAP values are now given by the solution to Equation \ref{eq:SHAPhpi} that satisfies
\begin{equation}
\label{eq:SHAPcondition}
f_x(v) = E[f(v)|v_S],
\end{equation}
where $S$ represent the set of non-zero indices in $v$. This constraint ensures that the SHAP values do not violate the consistency and/or the local accuracy properties, for more information see \cite{Lundberg_2017}. Thus, in the end, we get a specific contribution of each explanatory variable to the prediction \added{of the credit rating} for every individual observation considered. As the OL model allows for easy interpretation through its coefficients, this method is \deleted{only }used for the MLP\added{,}\deleted{and} CART\added{, SVM, and NB}. For the SHAP values Python's SHAP package is used \citep{Lundberg_2017}.

\subsection{Model evaluation}
\label{sec:modelevaluation}
Following common practice in literature \citep{Ozturk_2015, Reusens_2017}, for each model, we determine what percentage of the predictions was exactly right, 1 or 2 notch(es) too high and 1 or 2 notch(es) too low. Where a credit rating prediction is said to be $u$ notch(es) too low (high) if the predicted class is the $u$ class(es) below (above) the actual rating class.  \\
\indent Predictions are made using random split 10-fold cross-validation. That is, the data is split into 10 approximately equal subsets of which 9 are used to train the model, and the subset that was left out is used for evaluation of the out-of-sample predictive accuracy. By rotating the 10-folds, we obtain the out-of-sample accuracy of the model on the entire data set. We use the averages of 100 replications of this procedure, each time using different 10-fold data splits, thus making sure that results are not dependent on one specific random split. \added{Note that for every iteration of cross-validation the hyperparameters are the same. However, since a new training set requires retraining of the model, the parameters are updated in every iteration.}\\
\indent \added{Next to random cross-validation, we can investigate the performance of the different techniques when certain years are left out entirely. That is, all the observations of one year are either assigned to the training or the test set. This allows us to see if there are year-specific dependencies that the algorithms pick up on and use in their predictions.}

\section{Data}
\label{sec:data}

\begin{figure}
	\includegraphics[width=1\textwidth]{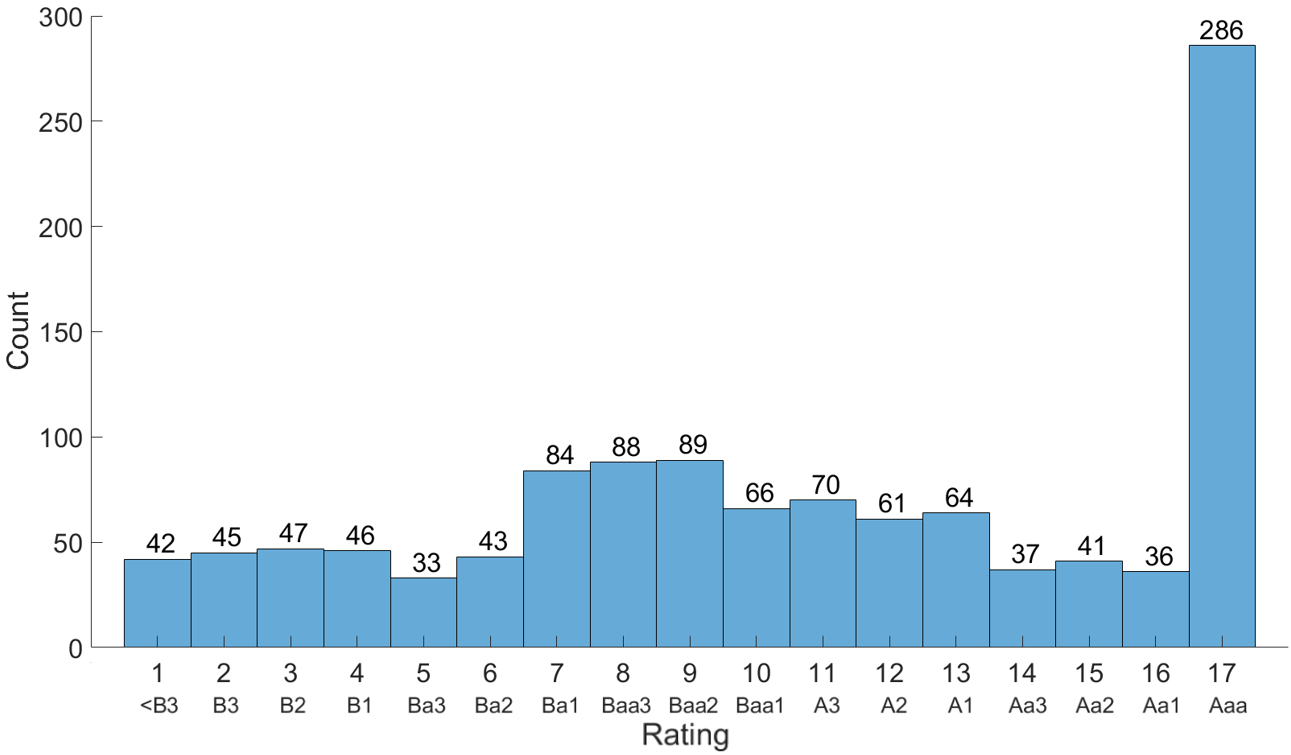}
	\caption{Histogram of ratings given by Moody's with numerical conversion for the period 2001-2019, ratings as of January 1 of each year are used. Full list of countries included can be found in Appendix \ref{app:listofcountries}. 1178 observations in total.}
	\label{fig:Histogram}
\end{figure}

\noindent We use Moody's' sovereign credit ratings for a variety of 62 developed and developing countries, among which Brazil, Canada, Morocco and Thailand, from 2001 to 2019.\footnote{Obtained from countryeconomy.com.} A histogram of the ratings with their alphabetical and numerical rating is shown in Figure \ref{fig:Histogram}. In this figure, we see that the data set contains a good mixture of the different categories, however, class 17 (Aaa) is, with 286 observations, significantly overrepresented. This is due to the fact that most of the Aaa countries stayed in this category throughout the entire period. There is therefore a trade-off between having enough different countries with an Aaa rating to train on and making sure the share of Aaa ratings does not become too large. The share of other ratings seems to be well balanced, with numeric ratings 7, 8 and 9 appearing a little more often (approximately 85 times). The full list of countries can be found in Appendix \ref{app:listofcountries}, and the transformation from Moody's to numerical ratings in Appendix \ref{app:ratingtransformations}.\\
\indent Previous research investigating the determining factors of sovereign credit ratings, mostly using linear regressions or an Ordered Logit/Probit model, shows that there are factors that frequently prove to be important. \cite{Cantor_1996} found that GDP per capita, GDP growth, inflation, external debt, level of economic development, and default history are determining factors in the credit rating process. The importance of these factors is also found by other researchers \citep{Afonso_2011,Ferri_1999,Gaillard_2009,Dimitrakopoulos_2016}, although there are some contrasting outcomes, for example, \cite{Ferri_1999} finds GDP per capita to be unimportant. Apart from economic and fiscal indicators, governance indicators prove to be important in the credit rating process, as was shown by \cite{Ozturk_2014}. Other factors that also commonly proved to be influential in these studies are: government balance, current account balance, and government effectiveness. \added{We include all these variables in this study to make sure the variables which proved to be important in previous studies are taken into account. One exception is default history, which is not included due to the low occurrence of sovereign defaults. Training such a factor would be very difficult on a few defaults, and results could therefore be spurious.} \deleted{We have included all these factors in our study, except for default history, because very few sovereign defaults have occurred during the period and estimating its effect would thus be very difficult.}\\
\indent In accordance with the above mentioned literature, we use a combination of economic figures, fiscal indicators, and governance indices as explanatory variables. These variables are: unemployment rate\footnotemark (measured in \%, with expected sign --), government balance\footnotemark[\value{footnote}] (\% of GDP, +), current account balance\footnotemark[\value{footnote}] (\% of GDP, +), inflation as measured by CPI\footnotemark[\value{footnote}] (\%, --), GDP per capita\footnotemark[\value{footnote}] (\$, +), government debt\footnote[\value{footnote}]{Obtained from the International Monetary Fund.} (\% of GDP, --), GDP growth\footnotemark (annual \%, +), regulatory quality index\footnotemark[\value{footnote}] (+), and political stability and absence of violence/terrorism index\footnote[\value{footnote}]{Obtained from the World Bank.} (+). Where the variables up to and including GDP growth are economic \& fiscal indicators, while the latter two are measures of governance. The regulatory quality index measures perceptions of the government's ability to formulate and implement policies and regulations that permit and promote private sector development. While the political stability and absence of violence/terrorism index captures the perceptions of likelihood of political instability and/or politically-motivated violence. Both these variables have values ranging from approximately -2.5 to 2.5, where a higher score indicates better regulatory quality or higher political stability. The values of all the explanatory variables for year $t$ are used to model Moody's sovereign credit ratings as of January 1 of year $t+1$. \\ 
\indent This set of explanatory variables represents three main factors in the credit rating process: the strength of the economy, the level of debt, and the willingness to repay. A strong economy is expected to be better capable of repaying its debt and preventing the debt burden to get out of control. Typical for strong economies are: a low unemployment rate, low (though not negative) and stable inflation, a high GDP per capita, and a high GDP growth. The debt position of a country is given by the government debt, and the government balance shows if the total debt sum (in \$) is increasing or decreasing. Finally, regulatory quality and political stability can give an indication of the willingness of a country to repay their debt, but also of the economic climate for the private sector in a country.\\
\indent The descriptive statistics for the explanatory variables are shown in Table \ref{tab:Descriptivemacro}. This table reports the median, mean, standard deviation, and 1\% \& 99\% percentiles of all variables. We immediately observe that the average country experienced an economic expansion during this period, as can be seen from the positive average GDP growth of 3.2\%. Furthermore, we observe that the average unemployment rate for the period is 7.9\%, but very low unemployment rates of below 2.0\% are also observed, for example for Thailand and Singapore. The average government debt is 55.0\% for the period, with a small number of countries, such as Japan and Venezuela, having a debt of over 180\%, which can be considered extremely large. \added{Inflation contains some outliers, which is due to Venezuela in 2015-2018.}\footnote{\added{To check the sensitivity of our findings to these outliers, the OL model (expected to be most heavily influenced by outliers), is also estimated without Venezuela. While the coefficient of inflation changes a bit (coefficient is negative in both cases), cross-validated accuracy remains unchanged at 33\%.}} Lastly, the gap between the very rich and very poor countries is large, with some of the rich countries (Luxembourg, Norway) having a GDP per capita that is up to 80 times higher than some of the poor countries (Honduras, Pakistan). \added{Correlations between the variables mentioned in this section are shown in Appendix \ref{app:correlations}.} 

\begin{table}
	\caption{Descriptive statistics of the macroeconomic and socioeconomic variables for the period 2000-2018, 1178 observations. Data obtained from the IMF and the World Bank. Full list of countries included can be found in Appendix \ref{app:listofcountries}.}
	\label{tab:Descriptivemacro}
	\begin{tabular}{@{}lrrrrr@{}}
		\hline\noalign{\smallskip}
		Variable & Med. & Mean & Std    & 1\%    & 99\%    \\
		\noalign{\smallskip}\hline\noalign{\smallskip}
		GDP growth (\%)             & 3.2    & 3.2  & 3.4    & -7.4  & 10.9    \\
		Inflation (\%)          & 2.7    & 60.1 & 1904.7 & -1.1   & 31.1 \\
		Unemployment rate (\%)      & 6.9    & 7.9  & 4.6    & 1.5   & 25.2   \\
		Current acc. (\% of GDP) & -0.4   & 9.6  & 50.2   & -91.0 & 236.5   \\
		Gov. balance (\% of GDP) & -2.2   & -1.9 & 4.2    & -11.6  & 12.0    \\
		Gov. debt (\% of GDP) & 46.9   & 55.0 & 33.8   & 7.4    & 182.6 \\
		Political stability        & 0.4    & 0.3  & 0.9    & -2.3   & 1.6     \\
		Regulatory quality         & 0.8    & 0.7  & 0.8    & -1.3   & 2.0     \\ 
		GDP per capita (1000\$)     & 13.5 & 22.1 & 22.0   & 0.9    & 101.8 \\
		\noalign{\smallskip}\hline
	\end{tabular}
\end{table}

\section{Results}
\label{sec:results}

In this section, we present the results obtained in this study. First, we discuss the accuracies of the different models when evaluated using cross-validation. Second, the determining factors for each model are analysed individually, and compared to those of the other models.

\subsection{Cross-validated accuracy}
\label{sec:results:accuracy}
The accuracies of the MLP, CART, \added{SVM, NB,} and OL, determined using 100 replications of 10-fold cross-validation, are shown in Table \ref{tab:accuracy}. In this table, for every model, we present the percentage of predictions exactly right, 1 or 2 notch(es) too high or too low, the number of predictions correct within 1 and 2 notch(es), and the Mean Absolute Error (MAE). \added{The random split shows how accurate the models are on a purely random split, while the year-based split forces an entire year of observations to be either in the training or in the test set.}\\
\indent \deleted{The}\added{Based on random split cross-validation} MLP performs best with an accuracy of 68.3\%, and 85.7\% of predictions correct within 1 notch. MLP outperforms CART\added{, SVM, NB,} and OL, with respective accuracies of 58.6\%\added{, 41.4\%, 37.6\%,} and 33.1\%, significantly on a 99\% significance level.\deleted{CART outperforms the OL model significantly and is, based on performance, much closer to the MLP than to the OL model.} \added{CART outperforms the other models significantly and is, based on performance much closer to MLP than to SVM. Both SVM and NB underperform compared to the other ML techniques, but, nonetheless, still outperform the OL model.} These results confirm earlier findings that \deleted{MLP and CART}\added{Machine Learning methods} outperform linear models based on accuracy, see, for example, \cite{Bennell_2006,DeMoor_2018,Ozturk_2015,Ozturk_2016}.\\
\indent A nice symmetry in over- and underrating is observed for all models. This shows that none of the models has a tendency to consistently rate higher or lower than Moody's. Additional related results, available upon request, show that no country is persistently under- or overrated by MLP and CART. \added{SVM, NB, and} OL, on the other hand, have that tendency\added{, and sometimes consistently give too high or too low ratings for certain countries.} \deleted{and, for example, continuously underrates France and Belgium, and overrates Bulgaria and Cyprus compared to Moody's.}\added{A misclassification analysis for the best performing method (MLP) is presented in Appendix \ref{app:Misclassanalysis}. The table shows that, next to the relatively high amount of correct classifications, large deviations are rare. Often when the MLP misclassifies it predicts one notch too high or too low, although there are some cases where the difference is large. Also noticable is the fact that the correct predictions are well spread across the different categories.}.\\
\indent \added{There are multiple possible causes for the relatively large differences in accuracy between the different techniques. MLP and CART have proven to be well suited for sovereign credit rating and were therefore expected to perform well, while for SVM and NB previous results were mixed \citep{Ozturk_2015,Ozturk_2016}. This is, in the case of SVM and NB, not unexpected since the performance of these techniques can be very dependent on the data set, see, for example, \cite{Hastie_2009,Rish_2001}. The outperformance of OL by all ML techniques is also sensible, and there are multiple possible causes.}\deleted{There are multiple possible causes for the relatively large difference in accuracy between the ML techniques and the OL model.} First, ML techniques are able to pick up on non-linear relations, where the OL model with its assumption of linear relations cannot. Research has shown that there are non-linear effects in the sovereign credit rating process, so assuming linear relations is likely to harm performance \citep{Reusens_2016}. Second, the ML techniques have more modelling freedom to pick up on subjective factors of the CRAs, which \cite{DeMoor_2018} show to be especially large for low-rated countries.\\ 
\indent \added{The year-based cross-validation shows results very similar to that of random split cross-validation. In fact, the ranking of the methods is identical. The consistency of the results for random and year-based splits provides an indication that there are no year-specific effects, such as caused by the state of the world economy, that the algorithms pick up on and use in their predictions for other countries in the same year.
}

\begin{table}
	\caption{Averages of 100 replications of 10-fold cross-validated predictions for MLP, CART, SVM, NB, and OL. All numbers, except for MAE, given in \%.}
	\label{tab:accuracy}       
	\begin{tabular}{l@{\hspace{7mm}}cccccccc}
		\hline\noalign{\smallskip}
		& \multicolumn{7}{c}{Correct prediction percentage} &\\
		\cline{2-8}\noalign{\smallskip}
		& 2 below & 1 below & Exact & 1 above & 2 above & Within 1 & Within 2 & MAE \\
		\noalign{\smallskip}\hline\noalign{\smallskip}
		\multicolumn{9}{l}{Random split}\\
		MLP  & 3.9 & 8.4  & 68.3 & 9.0  & 3.6  & 85.7 & 93.2 & 0.64 \\
		CART & 5.6 & 8.7  & 58.6 & 9.1  & 5.2  & 76.4 & 87.2 & 1.00 \\ 
		SVM & 7.2 & 8.5  & 41.4 & 8.7  & 6.4  & 58.7 & 72.3 & 1.89 \\ 
		NB & 8.0 & 10.6  & 37.6 & 12.7  & 8.4  & 60.9 & 77.3 & 1.57 \\ 
		OL   & 9.8     & 10.3    & 33.1  & 13.2    & 10.3    & 56.6     & 76.7     & 1.60 \\
		\noalign{\vskip 3mm}
		\multicolumn{9}{l}{Year-based split}\\
		MLP  & 3.6 & 8.5  & 68.6 & 9.1  & 3.5  & 86.1 & 93.2 & 0.65 \\
		CART & 5.4 & 8.9  & 58.9 & 9.5  & 4.9  & 77.3 & 87.6 & 0.98 \\
		SVM & 7.3 & 8.5  & 41.6 & 9.0  & 6.3  & 59.1 & 72.7 & 1.86 \\ 
		NB & 8.0 & 10.6  & 37.8 & 12.6  & 7.7  & 61.0 & 76.8 & 1.59 \\ 
		OP   & 9.4     & 11.2    & 32.3  & 12.8    & 10.0    & 56.4     & 75.7     & 1.64 \\
		\noalign{\smallskip}\hline
	\end{tabular}
\end{table}

\subsection{Determining factors} 
\label{sec:results:detfactors}

In order to get an insight into the sovereign credit ratings, we analyse the determining factors for every model. We obtain the determining factors of the MLP\added{,}\deleted{and} CART\added{, SVM, and NB} by using SHAP values, as discussed in Section \ref{sec:SHAPvalues}, and those of the OL model by looking at each variables' coefficients and their significance. 

\begin{figure}
	\centering
	\begin{subfigure}[MLP]{
		\includegraphics[width=.45\textwidth]{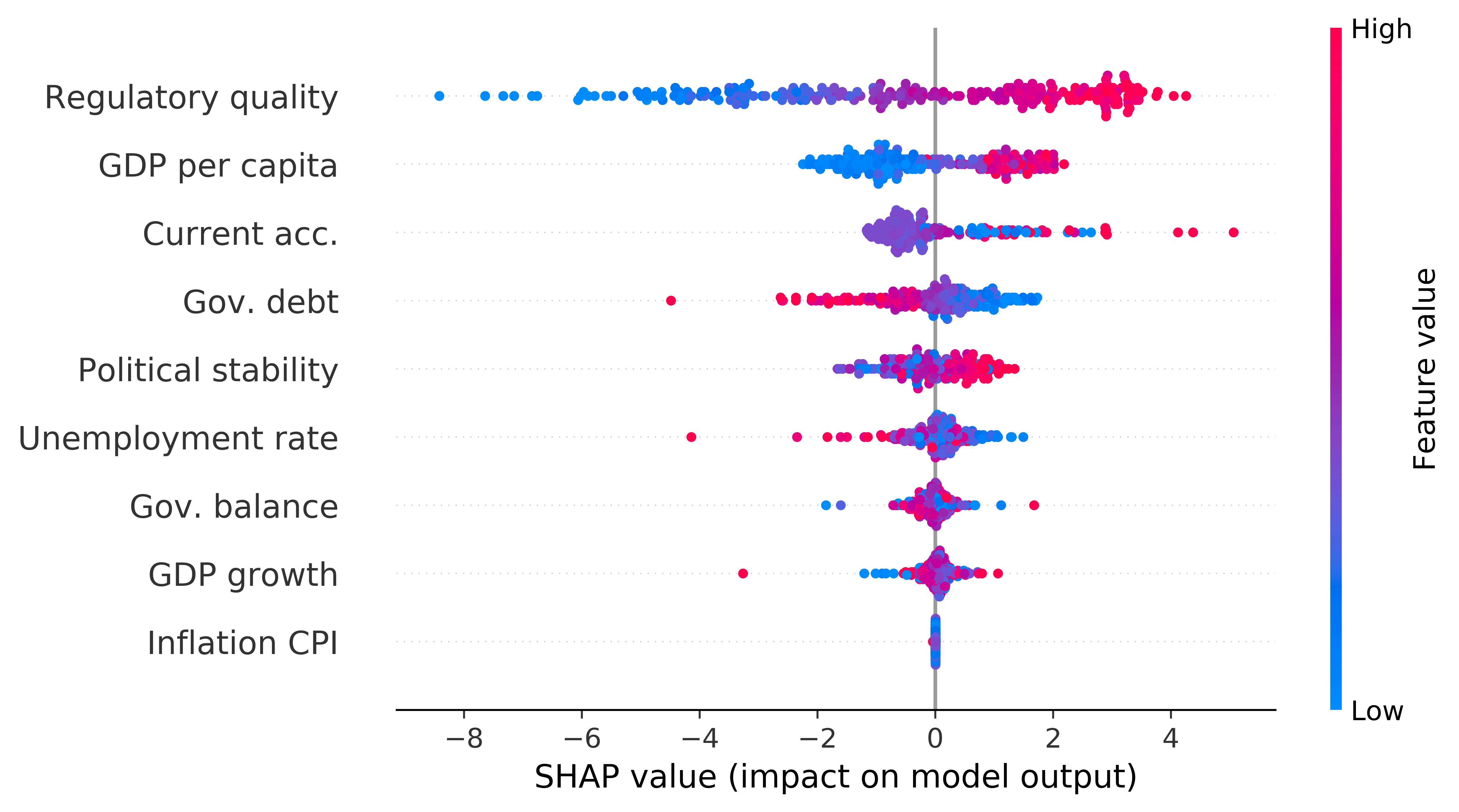}}
	\end{subfigure}
	~ 
	\begin{subfigure}[CART]{
		\includegraphics[width=.45\textwidth]{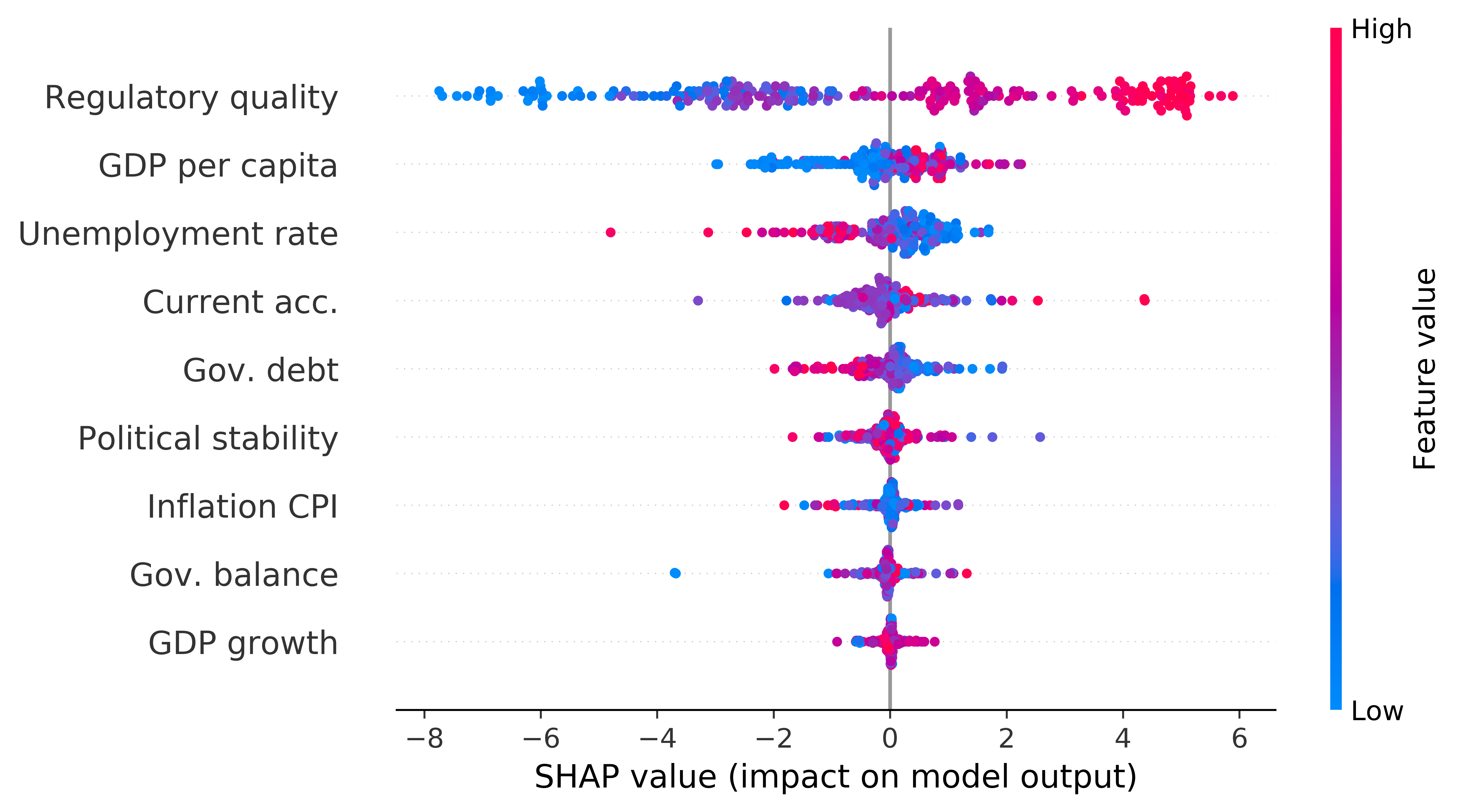}}
	\end{subfigure}
	{\vskip 5mm}
	\begin{subfigure}[SVM]{
		\includegraphics[width=.45\textwidth]{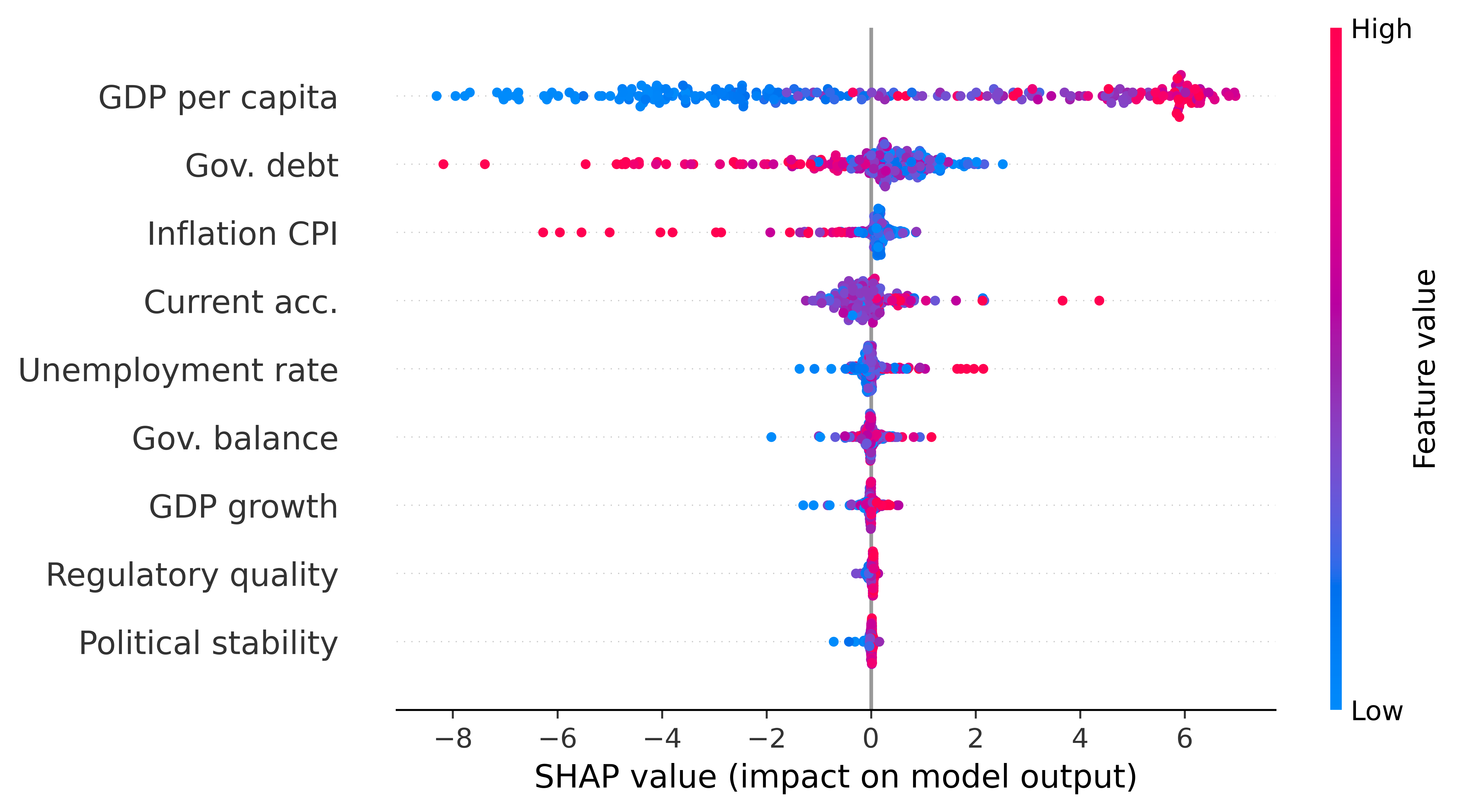}}
	\end{subfigure}
	~ 
	\begin{subfigure}[NB]{
		\includegraphics[width=.45\textwidth]{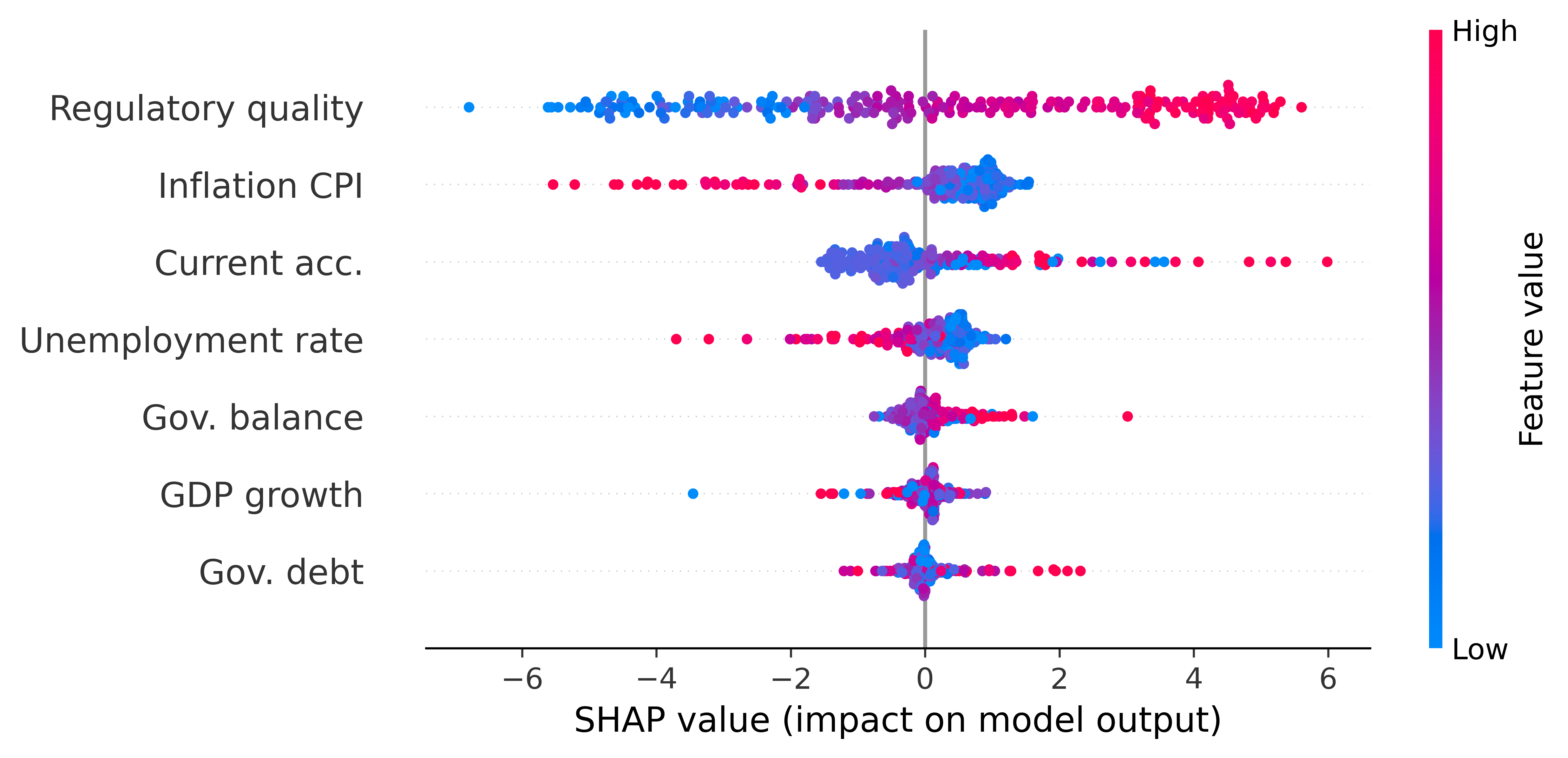}}
	\end{subfigure}
	\caption{SHAP values plots for the MLP, CART, SVM, and NB, explanatory variables are ranked from highest mean absolute SHAP value (regulatory quality in the case of MLP) to lowest (inflation in the case of MLP). Individual dots represent data points that have been evaluated, where the color indicates whether the value is relatively high (red) or low (blue) for that explanatory variable. The x-axis shows the impact of the particular feature on the prediction, i.e. the number of notches the prediction deviates from the baseline prediction when that feature is included.}
	\label{fig:SHAPfigures}
\end{figure}

\paragraph{Multilayer Perceptron} \mbox{} \\
SHAP values are calculated for every variable used in the MLP to isolate their effects, and are shown in Figure \ref{fig:SHAPfigures}. We immediately observe clear patterns for the regulatory quality and GDP per capita, the most important and second most important variable respectively. A higher value for either variable\added{, indicated in red in the figure,} is associated with an increase in the credit rating, which is in line with economic theory. The importance of regulatory quality is perhaps surprising, since one would expect financial indicators to be most important in an assessment of credit risk. However, regulatory quality might be the best indicator of the economic climate for the private sector in a country, which in turn might be the most relevant factor in separating creditworthy from non-creditworthy countries. Furthermore, regulatory quality also gives an indication of the willingness to pay. That GDP per capita turns out to be an important factor in the credit rating process is not unexpected, since it is a good measure of the relative size of the economy and wealth of a country, and has proven to be important in previous studies \citep{Bissoondoyal_2005, Gaillard_2009, Afonso_2011}.\\
\indent The next variable, current account balance, shows a positive influence on the credit rating when the value is either relatively low or relatively high, and a negative influence on the credit rating for an average value. This non-linear relation is also visible in the data, as stronger economies are more towards the extremes. The Netherlands and Germany for example have a very high current account balance, while that of the United Kingdom and Australia is very low. Current account balance is directly followed by government debt, where a higher debt is associated with a lower rating, which is in line with economic intuition. Political stability and unemployment rate, ranking 5$^{\textrm{th}}$ and 6$^{\textrm{th}}$, also show a pattern although less pronounced than the previously discussed variables. Here, a higher political stability and/or a lower unemployment rate are associated with an increase in the credit rating, and vice versa.\\
\indent The three least important variables, being government balance, GDP growth, and inflation, show no clear effect. Inflation even seems to have no influence at all. The relative unimportance of these three factors is quite sensible. A negative government balance is generally a bad sign, because it increases the government debt. However, as previously discussed, a higher rating leads to a lower interest rate and therefore less inclination to keep the debt low. Government balance is thus not such a helpful factor in the credit rating process. GDP growth and inflation do not lend themselves very well for distinction between creditworthy and non-creditworthy countries. In the case of GDP growth, we observe that lower rated countries have on average a higher GDP growth, but a lower cumulative GDP growth over long periods, which in the end determines the long-term growth of the economy. While hyperinflation is obviously a bad sign, and should lead to a low rating, inflation offers no clear guidance for the other values.

\paragraph{Classification and Regression Trees}\mbox{} \\

\noindent The same procedure as for the MLP is repeated to isolate the influence of variables in the CART, the plot containing SHAP values for the CART is shown in Figure \ref{fig:SHAPfigures}. Furthermore, to facilitate comparison with the MLP, Table \ref{tab:Comparingdetfactors} shows the explanatory variables ranked on importance for every model. In the CART, similar to the MLP, regulatory quality and GDP per capita are the most important and second most important variables respectively. As expected, a higher regulatory quality and a higher GDP per capita  are both associated with a higher credit rating.\\
\indent In the CART, just as in the MLP, unemployment rate, current account balance, government debt, and political stability rank 3$^{\textrm{rd}}$ to 6$^{\textrm{th}}$, although the exact order differs. The unemployment rate shows a clear relation to the credit rating, where a higher unemployment rate is associated with a lower credit rating. The 4$^{\textrm{th}}$ most important variable, current account balance, shows the same non-linear behaviour as the MLP, with average values resulting in a lower credit rating, and the extremes in a higher one. However, in this case, the effect is much less pronounced than in the MLP. Government debt proves to have a negative effect on the credit rating, which is in line with economic theory. In contrast to the MLP, political stability shows no clear relation to the credit rating in the CART.\\
\indent The three least important variables in the CART match those of the MLP. Inflation, government balance, and GDP growth show no distinct relation to the credit rating.

\begin{table}
	\caption{Ranking of the variables based on influence in the predictions of MLP, CART, \added{SVM, NB,} and OL. The higher the rank, the more important the variable, with 1 being the most influential variable. \added{NB and OL do not make use of all variables, as leaving out some variables improves their out-of-sample predictive accuracy. These variables therefore receive no ranking.} \deleted{As gov. debt is excluded from the OL model, no rank can be assigned to the variable, however, this could be interpreted as ranking last.} }
	\label{tab:Comparingdetfactors}
	\begin{tabular}{@{}l@{\hspace{7mm}}ccccc@{}}
		\hline\noalign{\smallskip}
		& MLP & CART & SVM & NB & OL \\ 
		\noalign{\smallskip}\hline\noalign{\smallskip}
		Regulatory quality  & 1   & 1 & 8 & 1 & 1  \\
		GDP per capita       & 2   & 2 & 1 & - & 5  \\
		Current acc.         & 3   & 4 & 4 & 3  & 7  \\
		Gov. debt            & 4   & 5 & 2 & 7  & -  \\
		Political stability & 5   & 6 & 9 & -  & 6  \\
		Unemployment rate    & 6   & 3 & 5 & 4   & 2  \\
		Gov. balance         & 7   & 8 & 6 & 5  & 8  \\
		GDP growth           & 8   & 9 & 7 & 6 & 3  \\
		Inflation CPI        & 9   & 7 & 3 & 2  & 4  \\
		\noalign{\smallskip}\hline\noalign{\smallskip}
		Predictive accuracy & 68\% & 59\% & 41\% & 38\% & 33\% \\
		\noalign{\smallskip}\hline
	\end{tabular}
\end{table}

\paragraph{Support Vector Machines}\mbox{} \\
\added{The investigation of explanatory variables using SHAP values for SVM is also shown in Figure \ref{fig:SHAPfigures}. Where the determining factors in MLP and CART were fairly similar, for SVM they differ substantially. GDP per capita ranks first here, which also ranks high in the MLP and CART. Thereafter comes government debt, which ranked lower in the other models, but proves to be important for the SVM. The effects of these variables are as expected with a higher GDP per capita and/or a lower government debt associated with a higher credit rating.\\
\indent Inflation, current account balance, and unemployment rate make up rank 3 to 5 in the SVM and have relationships that match economic intuition, or in the case of current account balance match the influence as found in the other models. For the other variables, government balance, GDP growth, regulatory quality, and political stability, no distinct relationship between explanatory variable and credit rating is found. This is especially surprising for regulatory quality, which proved very influential in the other models.\\
\indent The large difference between the determining factors for SVM versus MLP and CART likely stems from the way in which SVM handles the variables. As SVM uses a kernel (RBF in this case), it uses a non-linear transformation of the explanatory variables as input for the model. While this may increase cross-validated accuracy, it makes interpretation of the variables more difficult and results less predictable. We believe that, due to the more difficult interpretation of the variables, combined with the relatively low accuracy compared to MLP and CART, the results of the SVM should not be weighted heavily in the interpretation of the determining factors for sovereign credit ratings.}

\paragraph{Naïve Bayes}\mbox{} \\
\added{Determining factors for the NB model are also shown in Figure \ref{fig:SHAPfigures}. Compared to the other models NB misses two variables, GDP per capita and political stability, this is due to the fact the removing them improved cross-validated accuracy as they are highly correlated with regulatory quality (see Appendix \ref{app:correlations}). That does not necessarily mean that these variables are not relevant in the credit rating process, however, since both are correlated with regulatory quality, they violate NB's assumption of independence.\\
\indent Regulatory quality again proves to be the most important variable, where, conform economic intuition, a higher regulatory quality is associated with a higher credit rating. Inflation ranks second in the NB, a higher inflation results in a lower credit rating, as was to be expected. The high ranking of inflation is in contrast to MLP and CART, where inflation ranks very low, the difference here is likely due to dependencies between the variables which the NB does not pick up on.\\
\indent The next two variables, current account balance, and unemployment rate, also rank fairly high in the other models, and the effect they have in the NB model matches the influence in the other models. That is, average values for current account balance are associated with a lower credit rating, while the extremes relate to a higher one, and a higher unemployment rate results in a lower credit rating. The last three variables, government balance, GDP growth, and government debt, show no clear relationship with the credit rating in the NB model. In the case of government debt this is strange, as it proved important (ranking 3$^{\textrm{rd}}$ to 4$^{\textrm{th}}$) in the other models. The low ranking of government debt is likely to be due to the independence assumptions, whereas government debt is only useful in combination with other variables, as further discussed in the next section.
}

\paragraph{Ordered Logit}\mbox{} \\
Extracting the determining factor of the OL model is relatively simple, as the model only uses linear relations. The significance of the coefficients of different variables, combined with the sign, tells us how important a variable is and if the relation to the credit rating is positive or negative. The estimated coefficients, together with their standard error and $p$-value, are shown in Table \ref{tab:OPcoefficients}. No coefficient for government debt is estimated because excluding government debt from the model results in a higher cross-validated accuracy. The ranking of importance for all the variables in the OL model is also shown in Table \ref{tab:Comparingdetfactors}, together with those of the other models. \\
\indent Again, regulatory quality proves to be the most important variable, where the positive sign of the coefficient shows that the relation is positive, as was the case for the MLP\added{, NB,} and CART. That regulatory quality is most important in all models \added{except SVM} is a strong indication that it is a very important factor in the credit rating process. The second most important variable in the OL model is unemployment rate, where a higher unemployment rate is associated with a lower rating. The unemployment rate is followed by GDP growth and inflation, which ranked very low in the \deleted{other }two \added{best performing} models, and in the case of GDP growth, the sign is counter to expectation. However, as previously discussed, lower rated countries have a higher GDP growth on average, just not a higher cumulative growth. The OL model, with its linear relations, therefore finds a negative relation between GDP growth and the credit rating \added{as was also found in the study of \cite{Ozturk_2016}}. GDP per capita ranks 5$^{\textrm{th}}$ in the OL model, where it ranked 2$^{\textrm{nd}}$ in the \deleted{other two models}\added{MLP and CART, and 1$^{\textrm{st}}$ in the SVM}. The sign does match expectations with a higher GDP per capita associated with a higher credit rating. The next variable is political stability, \deleted{here the three models more or less agree on the importance of the factor. However,}\added{where} the sign of the coefficient in the OL model is counter to economic theory, which could be an effect of the inclusion of a lot of variables in a linear setting\added{, and the high correlation with regulatory quality}\footnote{\added{Only using political stability as explanatory variable in the OL results in a positive coefficient, while using regulatory quality and political stability as explanatory variables gives a positive coefficient for regulatory quality and a negative for political stability.}}. The last two variables, current account balance and government balance have a positive influence on the credit rating, which is in line with economic theory\added{, and with results of \cite{Ozturk_2016}}. That current account balance is relatively unimportant in the OL models compared to the other models makes sense, as the OL model cannot pick up on the non-linear relation that the other models find.\\
\indent The rank of government balance in the OL model is similar to that of the other \deleted{two }models. While the coefficients of current account balance and government balance are insignificant, they do contribute to the cross-validated accuracy and are therefore included in the model. The only factor that does not contribute to a higher cross-validated accuracy is government debt. This is strange, since government debt is commonly assumed to be a very important factor in the creditworthiness of a country. Countries that already have a lot of debt might be less able to repay new debt. It is nonetheless not a clear distinguishing factor on its own. There are also Aaa rated countries that have a lot of debt, since they have less inclination to keep the debt low due to the low interest rates that they pay. Government debt therefore seems to only be influential when taking into account other factors at the same time, and is thus not useful for the OL model. These results confirm earlier findings that government debt is a useful variable to split data on, but not necessarily useful in a regression model, see, for example, \cite{Bozic_2013,Reusens_2016}.\\
\indent The large similarities in determining factors, especially between \added{the two best performing models,} MLP and CART, are surprising, since the modelling techniques are quite different. This makes it more likely that some of variables found to be important in this study, such as regulatory quality, have a large influence on the credit rating.

\begin{table}
	\caption{Coefficients, standard errors and p-values for the Ordered Logit model. }
	\label{tab:OPcoefficients}
	\begin{tabular}{@{}lr@{\hspace{6mm}}rr@{}}
		\hline\noalign{\smallskip}
		& Coefficients & S.E. & p-value \\
		\noalign{\smallskip}\hline\noalign{\smallskip}
		GDP growth (\%)         & -0.0012      & 0.0000          & 0.0000  \\
		Inflation (\%)          & -0.0467      & 0.0118          & 0.0001  \\
		Unemployment rate (\%)  & -0.1082      & 0.0014          & 0.0000  \\
		Current acc. (\% of GDP)       & 0.0124       & 0.0182          & 0.4951  \\
		Gov. balance  (\% of GDP)      & 0.0409       & 0.1200          & 0.7331  \\
		Political stability & -0.2623      & 0.1567          & 0.0941  \\
		Regulatory quality  & 3.6001       & 0.0045          & 0.0000  \\
		GDP per capita (1000\$)    & 0.0337       & 0.0182          & 0.0642  \\
		\noalign{\smallskip}\hline
	\end{tabular}
\end{table}

\section{Conclusion}
\label{sec:concl}

\noindent This paper investigates the use of four Machine Learning techniques, Multilayer Perceptron (MLP)\deleted{ and}\added{,} Classification and Regression Trees (CART)\added{, Support Vector Machines (SVM), and Naïve Bayes (NB)}, \deleted{and}\added{next to} an Ordered Logit (OL) model, for prediction of sovereign credit ratings. MLP proves to be most suited for predicting Moody's ratings based on macroeconomic variables. Using random 10-fold cross-validation it reaches an accuracy of 68\%, and predicts 86\% of ratings correct within 1 notch. Thereby, it significantly outperforms \deleted{CART and OL with their respective accuracies of 59\% and 33\%}\added{CART, SVM, NB, and OL, with respective accuracies of 59\%, 41\%, 38\%, and 33\%}. \\ 
\indent Investigation of the determining factors, which has so far not been done for Machine Learning models in the sovereign credit rating setting, shows that there are common influential factors across the \added{best performing} models. Regulatory quality and GDP per capita are respectively the most important and second most important factor in the MLP and CART, with, as expected, a positive relation between both variables and the predicted credit rating. \deleted{This behaviour is also reflected by the signs of the respective coefficient in the OL model. Other, slightly less, influential variables are: current account balance, government debt, political stability and unemployment rate.} \added{While the ranking of variables in the MLP and CART is very similar, the other ML models sometimes deviate. Especially SVM, where regulatory quality ranks very low, differs in interpretation. This is likely due to the use of the RBF kernel, whereby it uses a non-linear transformation of the variable.} The behaviour of MLP and CART with respect to most \deleted{of these}variables is similar. A higher government debt and unemployment rate are associated with a lower credit rating, and for both models an average current account balance value leads to a lower rating while a relatively low or high value leads to a higher credit rating. The models differ on the interpretation of political stability. In the MLP, a higher value for political stability leads to a higher credit rating, but there is no clear relation in the CART. \deleted{Most of the previously mentioned effects are also observed in the signs of OL coefficients. However, the signs of GDP growth and political stability are in contrast to economic theory, where a higher GDP growth and/or political stability are associated with a lower credit rating, possibly due to inclusion of all variables jointly in the restrictive linear setting.}\\
\indent In short, we advice governments wanting to check their rating or investors deliberating an investment to use a MLP model, as this model proves to be most accurate. Sovereign credit ratings are heavily influenced by the regulatory quality and GDP per capita of a country. Expected changes in either of these factors could thus result in a credit rating change. Anticipating this possible change can be very valuable, as the credit rating has a major influence on the interest rate at which governments can issue new debt, and thus on the government budgets.\\
\indent We end this paper with a few recommendations for future research.\deleted{First, the determining factors of other Machine Learning techniques, such as Support Vector Machines (SVM), Naive Bayes (NB), and Bayes Net (BN), in the sovereign credit rating setting could be investigated.} \added{First, if, in the future, panel structures can be implemented in the Machine Learning algorithms, this would be very helpful in the modelling of sovereign credit ratings as it could further improve predictive accuracy of the ML models.} Second, \added{collecting and including}\deleted{the inclusion of} more explanatory variables might increase accuracy of some methods (most likely CART) and might lead to more insights into the relevant variables. \added{Third, an approach where variables are iteratively ommitted, as done in \cite{Ozturk_2014}, could give more insight into the interaction of the different variables.}

\bibliographystyle{apalike}  
\bibliography{Paper_sovereign_credit_ratings_arXiv_Update}   

\begin{thebibliography}{}

\bibitem[Afonso, 2003]{Afonso_2003}
Afonso, A. (2003).
\newblock Understanding the determinants of sovereign debt ratings: Evidence
  for the two leading agencies.
\newblock {\em Journal of Economics and Finance}, 27(1):56--74.

\bibitem[Afonso et~al., 2011]{Afonso_2011}
Afonso, A., Gomes, P., and Rother, P. (2011).
\newblock Short- and long-run determinants of sovereign debt credit ratings.
\newblock {\em International Journal of Finance \& Economics}, 16(1):1--15.

\bibitem[Baesens et~al., 2003]{Baessens_2003}
Baesens, B., Gestel, T.~V., Viaene, S., Stepanova, M., Suykens, J., and
  Vanthienen, J. (2003).
\newblock Benchmarking state-of-the-art classification algorithms for credit
  scoring.
\newblock {\em Journal of the Operational Research Society}, 54(6):627--635.

\bibitem[Bennell et~al., 2006]{Bennell_2006}
Bennell, J.~A., Crabbe, D., Thomas, S., and ap~Gwilym, O. (2006).
\newblock Modelling sovereign credit ratings: Neural networks versus ordered
  probit.
\newblock {\em Expert Systems with Applications}, 30(3):415 -- 425.
\newblock Intelligent Information Systems for Financial Engineering.

\bibitem[Bissoondoyal-Bheenick, 2005]{Bissoondoyal_2005}
Bissoondoyal-Bheenick, E. (2005).
\newblock An analysis of the determinants of sovereign ratings.
\newblock {\em Global Finance Journal}, 15(3):251 -- 280.
\newblock Special Issue.

\bibitem[Bozic and Magazzino, 2013]{Bozic_2013}
Bozic, V. and Magazzino, C. (2013).
\newblock Credit rating agencies: The importance of fundamentals in the
  assessment of sovereign ratings.
\newblock {\em Economic Analysis and Policy}, 43(2):157 -- 176.

\bibitem[Butler and Fauver, 2006]{Butler_2006}
Butler, A.~W. and Fauver, L. (2006).
\newblock Institutional environment and sovereign credit ratings.
\newblock {\em Financial Management}, 35(3):53--79.

\bibitem[Cantor and Packer, 1996]{Cantor_1996}
Cantor, R. and Packer, F. (1996).
\newblock Determinants and impact of sovereign credit ratings.
\newblock {\em Economic Policy Review}, 2(2).

\bibitem[Chollet et~al., 2015]{Keras}
Chollet, F. et~al. (2015).
\newblock Keras.
\newblock {https://keras.io}.

\bibitem[Dimitrakopoulos and Kolossiatis, 2016]{Dimitrakopoulos_2016}
Dimitrakopoulos, S. and Kolossiatis, M. (2016).
\newblock State dependence and stickiness of sovereign credit ratings: Evidence
  from a panel of countries.
\newblock {\em Journal of Applied Econometrics}, 31(6):1065--1082.

\bibitem[Elkhoury, 2009]{Elkhoury_2009}
Elkhoury, M. (2009).
\newblock Credit rating agencies and their potential impact on developing
  countries.
\newblock {\em UNCTD Compendium on Debt Sustainability}, pages 165--180.

\bibitem[Ferri et~al., 1999]{Ferri_1999}
Ferri, G., Liu, L.-G., and Stiglitz, J.~E. (1999).
\newblock The procyclical role of rating agencies: Evidence from the east asian
  crisis.
\newblock {\em Economic Notes}, 28(3):335--355.

\bibitem[Gaillard, 2009]{Gaillard_2009}
Gaillard, N. (2009).
\newblock The determinants of moody's sub-sovereign ratings.
\newblock {\em International Research Journal of Finance and Economics},
  31(1):194--209.

\bibitem[Hastie et~al., 2009]{Hastie_2009}
Hastie, T., Tibshirani, R., and Friedman, J. (2009).
\newblock {\em The Elements of Statistical Learning: Data Mining, Inference,
  and Prediction}.
\newblock Springer series in statistics. Springer.

\bibitem[Keskar et~al., 2016]{Keskar_2017}
Keskar, N.~S., Mudigere, D., Nocedal, J., Smelyanskiy, M., and Tang, P. T.~P.
  (2016).
\newblock On large-batch training for deep learning: Generalization gap and
  sharp minima.
\newblock {\em CoRR}, abs/1609.04836.

\bibitem[Kotsiantis et~al., 2006]{Kotsiantis_2006}
Kotsiantis, S.~B., Zaharakis, I.~D., and Pintelas, P.~E. (2006).
\newblock Machine learning: a review of classification and combining
  techniques.
\newblock {\em Artificial Intelligence Review}, 26(3):159--190.

\bibitem[Lessmann et~al., 2015]{Lessman_2015}
Lessmann, S., Baesens, B., Seow, H.-V., and Thomas, L.~C. (2015).
\newblock Benchmarking state-of-the-art classification algorithms for credit
  scoring: An update of research.
\newblock {\em European Journal of Operational Research}, 247(1):124 -- 136.

\bibitem[Liu et~al., 2011]{Liu_2011}
Liu, Q., Chen, C., Zhang, Y., and Hu, Z. (2011).
\newblock Feature selection for support vector machines with rbf kernel.
\newblock {\em Artificial Intelligence Review}, 36(2):99--115.

\bibitem[Luitel et~al., 2016]{Luiten_2016}
Luitel, P., Vanpée, R., and Moor, L.~D. (2016).
\newblock Pernicious effects: How the credit rating agencies disadvantage
  emerging markets.
\newblock {\em Research in International Business and Finance}, 38:286 -- 298.

\bibitem[Lundberg and Lee, 2017]{Lundberg_2017}
Lundberg, S.~M. and Lee, S.-I. (2017).
\newblock A unified approach to interpreting model predictions.
\newblock In {\em Advances in Neural Information Processing Systems 30}, pages
  4765--4774. Curran Associates, Inc.

\bibitem[Moor et~al., 2018]{DeMoor_2018}
Moor, L.~D., Luitel, P., Sercu, P., and Vanpée, R. (2018).
\newblock Subjectivity in sovereign credit ratings.
\newblock {\em Journal of Banking \& Finance}, 88:366 -- 392.

\bibitem[Ozturk, 2014]{Ozturk_2014}
Ozturk, H. (2014).
\newblock The origin of bias in sovereign credit ratings: Reconciling agency
  views with institutional quality.
\newblock {\em The Journal of Developing Areas}, 48(4):161--188.

\bibitem[Ozturk et~al., 2015]{Ozturk_2015}
Ozturk, H., Namli, E., and Erdal, H.~I. (2015).
\newblock Reducing overreliance on sovereign credit ratings: Which model serves
  better?
\newblock {\em Computational Economics}, 48:59 -- 81.

\bibitem[Ozturk et~al., 2016]{Ozturk_2016}
Ozturk, H., Namli, E., and Erdal, H.~I. (2016).
\newblock Modelling sovereign credit ratings: The accuracy of models in a
  heterogeneous sample.
\newblock {\em Economic Modelling}, 54:469 -- 478.

\bibitem[Panchal et~al., 2011]{Panchal_2011}
Panchal, G., Ganatra, A., Kosta, Y., and Panchal, D. (2011).
\newblock Behaviour analysis of multilayer perceptrons with multiple hidden
  neurons and hidden layers.
\newblock {\em International Journal of Computer Theory and Engineering},
  3(2):332--337.

\bibitem[Pedregosa et~al., 2011]{scikitlearn}
Pedregosa, F., Varoquaux, G., Gramfort, A., Michel, V., Thirion, B., Grisel,
  O., Blondel, M., Prettenhofer, P., Weiss, R., Dubourg, V., Vanderplas, J.,
  Passos, A., Cournapeau, D., Brucher, M., Perrot, M., and Duchesnay, E.
  (2011).
\newblock Scikit-learn: Machine learning in python.
\newblock {\em Journal of Machine Learning Research}, 12:2825--2830.

\bibitem[Pedregosa-Izquierdo, 2015]{mord}
Pedregosa-Izquierdo, F. (2015).
\newblock {\em {Feature extraction and supervised learning on fMRI : from
  practice to theory}}.
\newblock Theses, {Universit{\'e} Pierre et Marie Curie - Paris VI}.

\bibitem[Ramachandran et~al., 2017]{Rmachandran_2017}
Ramachandran, P., Zoph, B., and Le, Q.~V. (2017).
\newblock Searching for activation functions.
\newblock {\em CoRR}, abs/1710.05941.

\bibitem[Reusens and Croux, 2016]{Reusens_2016}
Reusens, P. and Croux, C. (2016).
\newblock Sovereign credit rating determinants: the impact of the european debt
  crisis.
\newblock {\em Available at SSRN 2777491}.

\bibitem[Reusens and Croux, 2017]{Reusens_2017}
Reusens, P. and Croux, C. (2017).
\newblock Sovereign credit rating determinants: A comparison before and after
  the european debt crisis.
\newblock {\em Journal of Banking \& Finance}, 77:108 -- 121.

\bibitem[Rish et~al., 2001]{Rish_2001}
Rish, I. et~al. (2001).
\newblock An empirical study of the naive bayes classifier.
\newblock In {\em IJCAI 2001 workshop on empirical methods in artificial
  intelligence}, volume~3, pages 41--46.

\bibitem[Shapley, 1953]{Shapley_1953}
Shapley, L.~S. (1953).
\newblock A value for n-person games.
\newblock {\em Contributions to the Theory of Games}, 2(28):307--317.

\end{thebibliography}
\appendix
\section{Appendix}
\subsection{MLP optimization}
\label{app:MLPoptimization}
There are basically two ways of optimizing model hyperparameters: grid search and Bayesian model-based optimization. While the Bayesian methods are more likely to give you the optimal setting, they give no insight into the different performance-complexity trade-offs. As that trade-off is important in this study, since interpretation suffers for more complex models, a grid search is used to optimize the MLP.\\
\indent In this grid search, five hyperparameters are optimized: number of hidden layers, number of neurons per hidden layer, dropout rate, number of epochs and batch size. The number of hidden layers and number of neurons per hidden layer, as previously explained, determine the structure of the MLP. The dropout rate gives the fraction of neurons that is dropped from the model at random. Randomly dropping neurons from the model prevents overfitting, as an overfitted model would perform very poorly when neurons are left out. The number of epochs and batch size determine how the internal model parameters are estimated. When a MLP is trained, it updates the parameters after working through a number of data points. That is, the internal parameters are not updated after evaluating every individual data point, but after evaluating a certain number of data points, a batch. A larger batch size thus means that the algorithm evaluates more data points before updating the parameters and vice versa for a small batch size. The number of epochs determines how many times the algorithm goes through the entire data set, especially for smaller data sets this number can be very large, often hundreds or thousands.\\
\indent Setting up a full grid, where all the different combinations are tested, is computationally extremely expensive, since the number of possible combinations becomes very large. We have therefore opted for two separate grid searches. First, one where the optimal structure is investigated: hidden layers, neurons and dropout rate. Thereafter, a second grid search in which the estimation of the optimal structure found in the first grid search is analysed: epochs and batch size.\\
\indent In the first grid search the following hyperparameters are considered: hidden layers [1, 2, 3], neurons [8, 16, 32, 64, 128, 256, 512] and dropout rate [0, 0.1, 0.2] using a batch size of 8 and 400 epochs. In general, one hidden layer suffices, only in cases where there are discontinuities in the data is more than one hidden layer required \citep{Panchal_2011}. Therefore, the grid search is limited to three hidden layers, to make sure that additional hidden layers do not improve performance. There are rules of thumb for selecting the number of neurons, such as that it should be between the size of the input and the size of the output layer. However, deviating from these rules often results in drastically improved performance. Even though 512 neurons seems excessively large, and is unlikely to improve performance compared to lower numbers, it is still evaluated to make sure no performance increase is obtained. The dropout hyperparameters are set in such a way that we can see if dropout is needed, or if dropping a significant, but not too large, fraction of the neurons improves performance. \\
\indent Thereafter, for the optimal structure, we investigate the estimation hyperparameters using: batch size [4, 8, 16, 32] and epochs [100, 200, 400, 800]. \cite{Keskar_2017} show that the batch size should be much smaller than the total number of data points in the set, and that using a large batch size decreases the ability of the model to generalize. For these reasons, we have decided to set an upper bound of 32 on the batch size. There are no clear guidelines for the optimal number of epochs, as this is highly dependent on the data set. The number of epochs is thus increased until performance of the MLP stops improving. If optimal performance in this grid is found at 800 epochs, the use of an even higher number of epochs is investigated.\\
\indent The results of the two grid searches are shown in Tables \ref{tab:MLPGridsearch} and \ref{tab:MLPGridsearchestimation}. The optimal performance-complexity trade-off is in our view given by the MLP with 1 hidden layer, 256 neurons, and a dropout rate of 0.1. Even though the MLP with 2 hidden layers, 256 neurons, and a dropout rate of 0.2 has a slightly higher accuracy, we deem the increase in accuracy too small to justify addition of a hidden layer. The results of the estimation grid search show that performance increases with more epochs, but levels off at about 200 epochs. Since we rather be on the safe side, we opted for 400 epochs. There is very little variation in performance between the different batch sizes, although combinations of a low number of epochs with large batches perform poorly. We have therefore decided to use a batch size of 8, for which the MLP was structure was optimized. 

\begin{table}
	\caption{MLP model structure optimization with hidden layers [1, 2, 3], neurons [8, 16, 32, 64, 128, 256, 512] and dropout [0, 0.1]. All models are estimated using batch size 8 and 400 epochs. Optimal structure, in terms of accuracy, consists of 2 hidden layer with 256 neurons with a dropout rate of 0.2. The best performance-complexity trade-off, underlined in the table, is obtained by the MLP with 1 hidden layer, 256 neurons and a dropout rate of 0.1. All numbers are given in \%.}
	\label{tab:MLPGridsearch}
	\begin{tabular}{@{}lr@{\hspace{7mm}}r@{\hspace{7mm}}r@{\hspace{7mm}}r@{\hspace{7mm}}r@{\hspace{7mm}}r@{\hspace{1mm}}r@{}}
		\hline\noalign{\smallskip}
		Batch size & 8  & \multicolumn{2}{c}{\multirow{2}{*}{No dropout}}  & \multicolumn{2}{c}{\multirow{2}{*}{Dropout 0.1}} & \multicolumn{2}{c}{\multirow{2}{*}{Dropout 0.2}}   \\
		Epochs & 400 & & & & & & \\
		\noalign{\smallskip}\hline\noalign{\smallskip}
		\multicolumn{1}{c}{Number of}  & \multicolumn{1}{c}{Neurons per} &\multicolumn{2}{c}{Accuracy}& \multicolumn{2}{c}{Accuracy}  &\multicolumn{2}{c}{Accuracy} \\
		\multicolumn{1}{c}{hidden layer(s)} & \multicolumn{1}{c}{hidden layer} & Mean   & Std   & Mean     & Std & Mean     & Std   \\
		\cline{3-8}
		\noalign{\smallskip}
		\multicolumn{1}{c}{\multirow{7}{*}{1}}         & 8   & 43.9              & 5.8         & 41.6               & 5.1          & 38.5               & 5.0          \\
		& 16  & 50.9              & 4.6         & 49.4               & 3.5          & 46.1               & 4.9          \\
		& 32  & 59.2              & 4.5         & 56.0               & 4.2          & 53.7               & 6.4          \\
		& 64  & 63.7              & 3.7         & 64.1               & 3.3          & 63.3               & 6.9          \\
		& 128 & 66.2              & 3.4         & 68.5               & 3.7          & 68.2               & 5.2          \\
		& 256 & 67.1              & 4.2         & \underline{69.7}               & 3.7          & 69.4               & 4.5          \\
		& 512 & 67.1              & 3.8         & 68.3               & 2.2          & 68.2               & 4.5          \\
		\noalign{\vskip 3mm}       
		\multicolumn{1}{c}{\multirow{7}{*}{2}}        & 8   & 40.8              & 5.3         & 38.8               & 4.7          & 37.0               & 5.0          \\
		& 16  & 48.7              & 4.1         & 43.7               & 4.3          & 41.3               & 4.7          \\
		& 32  & 55.6              & 4.6         & 56.6               & 4.4          & 51.9               & 5.3          \\
		& 64  & 62.1              & 4.0         & 64.3               & 4.2          & 65.7               & 6.0          \\
		& 128 & 65.5              & 4.5         & 68.9               & 4.4          & 68.3               & 4.3          \\
		& 256 & 67.7              & 2.8         & 69.5               & 3.1          & 70.0               & 4.2          \\
		& 512 & 68.8              & 2.2         & 68.1               & 1.9          & 69.6               & 2.9          \\
		\noalign{\vskip 3mm}
		\multicolumn{1}{c}{\multirow{7}{*}{3}}         & 8   & 39.5              & 8.3         & 38.5               & 5.9          & 34.0               & 4.8          \\
		& 16  & 45.6              & 5.4         & 45.0               & 5.8          & 38.9               & 4.8          \\
		& 32  & 52.8              & 3.0         & 54.7               & 5.2          & 46.3               & 5.8          \\
		& 64  & 62.8              & 5.1         & 64.8               & 3.8          & 63.7               & 4.6          \\
		& 128 & 65.5              & 4.5         & 67.9               & 4.0          & 69.6               & 4.8          \\
		& 256 & 66.4              & 2.5         & 68.1               & 3.9          & 68.4               & 4.9          \\
		& 512 & 68.1              & 1.8         & 68.3               & 4.1          & 66.9               & 4.9          \\ 
		\noalign{\smallskip}\hline 
	\end{tabular}
\end{table}

\begin{table}
	\centering
	\small
	\caption{MLP model estimation optimization with epochs [20, 50, 100, 200, 400, 800] and batch size [4, 8, 16, 32] on the MLP with 1 hidden layer, 256 neurons and dropout rate 0.1. Optimal estimation is achieved using 200 epochs and a batch size of 8.  All numbers are given in \%.}
	\label{tab:MLPGridsearchestimation}
	\begin{tabular}{lrrrr}
		\hline\noalign{\smallskip}
		Epochs & Batch size & Mean acc. & Std   \\ 
		\noalign{\smallskip}\hline\noalign{\smallskip}
		\multicolumn{1}{c}{\multirow{4}{*}{20}}      & 4          & 56.9      & 4.8 \\
		& 8          & 54.3      & 6.2 \\
		& 16         & 52.5      & 4.8 \\
		& 32         & 50.3      & 4.3 \\
		\noalign{\vskip 1mm}
		\multicolumn{1}{c}{\multirow{4}{*}{50}}    & 4          & 65.4      & 3.7 \\
		& 8          & 65.0      & 3.5 \\
		& 16         & 60.9      & 4.4 \\
		& 32         & 56.4      & 4.4 \\
		\noalign{\vskip 1mm}  
		\multicolumn{1}{c}{\multirow{4}{*}{100}}  & 4          & 67.7      & 4.1 \\
		& 8          & 67.5      & 5.9 \\
		& 16         & 65.4      & 4.1 \\
		& 32         & 63.2      & 4.7 \\
		\noalign{\vskip 1mm}
		\multicolumn{1}{c}{\multirow{4}{*}{200}}  & 4          & 67.5      & 3.8 \\
		& 8          & \underline{68.9}      & 5.2 \\
		& 16         & 67.1      & 3.7 \\
		& 32         & 67.6      & 3.3 \\
		\noalign{\vskip 1mm}
		\multicolumn{1}{c}{\multirow{4}{*}{400}}    & 4          & 68.0      & 4.8 \\
		& 8          & 68.7      & 5.1 \\
		& 16         & 68.0      & 4.1 \\
		& 32         & 67.2      & 5.6 \\
		\noalign{\vskip 1mm}
		\multicolumn{1}{c}{\multirow{4}{*}{800}}    & 4          & 68.3      & 4.8 \\
		& 8          & 68.2      & 5.1 \\
		& 16         & 68.2      & 5.4 \\
		& 32         & 68.2      & 4.1 \\ 
		\noalign{\smallskip}\hline 
	\end{tabular}
\end{table}

\subsection{CART Optimization}
\label{app:CARToptimization}
There are two ways in which a CART can be restricted: restricted growth and pruning. When restricting the growth of the CART, we limit the growth of the tree a priori, while with pruning we allow the tree to grow unobstructed but cut off branches afterwards.\\
\indent Limiting the growth of the CART can be done in multiple ways. In this study, we optimize the following settings: maximum depth, minimum samples for a split and minimum impurity decrease. Maximum depth limits the number of splits the tree is allowed to make by stopping after a certain depth is reached. That is, it limits the number of sequential splits the algorithm is allowed to make, counting from the root node. The minimum samples for a split restrict splitting, if the minimum number for a split is not reached, the algorithm is forced to make a leaf there. Lastly, a restriction can be set on the minimum impurity decrease, which means that the algorithm is only allowed to make a further split if that leads to a certain decrease in the Gini impurity (Equation \ref{eq:Gini}).\\
\indent For the CART, just as for the MLP, we use a grid search instead of Bayesian hyperparameter optimization techniques to get insight into the performance of the CART. Selecting the hyperparameter values to be included in the grid search requires some preliminary investigation, since the restrictions have to be adjusted to the size the tree would grow to when left unrestricted. Being too restrictive compared to unrestricted growth will significantly harm performance, whereas restrictions that do not restrict maximum growth have no effect at all. Initial investigation shows that a tree grown unrestrictedly on the full data set ends up with 320 leaves and a maximum depth of 20. Therefore, we have decided to use the following hyperparameter grid: maximum depth [10-20], minimum samples for a split [2, 3, 4, 5] and minimum impurity decrease [0-0.0002] in steps of 0.00001. \\
\indent Instead of limiting the growth of the tree, we can also prune it after letting it grow unrestricted. In this study, we make use of minimal cost-complexity pruning, that is, minimizing the cost-complexity criterion
\begin{equation}
\label{eq:pruning}
C_{\alpha}(L) = \sum_{l=1}^{|L|} \Bigg ( \sum_{u_i \subset R_l}^{}(y_i-\hat{y}_l)^2 \Bigg ) + \alpha|L|,
\end{equation}
\noindent where each $l$ represents a leaf and $|L|$ is the total number of leaves. The set $R_l$ contains all the data points in leaf $l$, $\hat{y}_l$ is the prediction for leaf $l$ and $\alpha$ is the factor punishing for complexity \citep{Hastie_2009}. In words, the cost-complexity criterion is the sum of squared errors with an additional factor that punishes for tree complexity in the form of the number of leaves. Thus, optimizing tree complexity is done by optimizing the factor $\alpha$.\\
\indent The results of the CART optimisation are very straight forward, any a priori limitation tree growth or pruning results in a decreases out-of-sample accuracy. An unrestricted CART is therefore most suited for sovereign credit rating predictions on this data set and is thus used in this research.

\subsection{SVM Optimization}
\label{app:SVMoptimization}
\added{
For the SVM, as was also done for MLP and CART, we use a grid search to find optimal hyperparameter settings. In this grid search we vary two hyperparameters: $C$ and $\gamma$. Of these, $C$ determines the cost of wrong classification, where a high $C$ leads to severe punishment of misclassifications, while a low $C$ allows the model to misclassify when determining the optimal hyperplane. The hyperparameter $\gamma$ determines how far the influence of a single training point reaches. When $\gamma$ is low, similarity regions are large, therefore, more points are grouped together, and vice versa for high $\gamma$ values. However, these hyperparameters also interact indirectly, if $\gamma$ is too large the model will overfit, irrespective of the value for $C$.\\
\indent In this grid search we use the following hyperparameter settings: $C$ [10, 1000, 100000] and $\gamma$ [10$^{-3}$, 10$^{-5}$, 10$^{-7}$]. Results are shown in Table \ref{tab:SVMgridsearch}. The highest accuracy is obtained by the SVM with $C$ equal to 100,000 and $\gamma$ at 10$^{-7}$.
}
\begin{table}
	\centering
	\small
	\caption{SVM grid search with $C$ [10, 1000, 100000] and $\gamma$ [10$^{-3}$, 10$^{-5}$, 10$^{-7}$]. Highest accuracy is obtained by the SVM with $C$ 100,000 and $\gamma$ 10$^{-7}$. All numbers are given in \%.}
	\label{tab:SVMgridsearch}
	\begin{tabular}{crrrr}
		\hline\noalign{\smallskip}
		$C$ & $\gamma$ & Mean acc. & Std   \\ 
		\noalign{\smallskip}\hline\noalign{\smallskip}
		\multicolumn{1}{c}{\multirow{3}{*}{10}}      & 10$^{-3}$         & 29.5      & 4.5 \\
		& 10$^{-5}$         & 28.9      & 2.0 \\
		& 10$^{-7}$        & 35.3      & 4.2 \\
		\noalign{\vskip 1mm}
		\multicolumn{1}{c}{\multirow{3}{*}{1000}}    & 10$^{-3}$         & 29.5      & 4.5 \\
		& 10$^{-5}$         & 28.7      & 2.2 \\
		& 10$^{-7}$        & 36.5      & 4.0 \\
		\noalign{\vskip 1mm}  
		\multicolumn{1}{c}{\multirow{3}{*}{100,000}}  & 10$^{-3}$         & 29.5      & 4.5 \\
		& 10$^{-5}$         & 29.2      & 2.6 \\
		& 10$^{-7}$        & 42.1      & 5.3 \\
		\noalign{\smallskip}\hline 
	\end{tabular}
\end{table}

\subsection{List of countries}
\label{app:listofcountries}
\noindent Countries included in the data set: Argentina, Australia, Austria, Belgium, Brazil, Bulgaria, Canada, China, Colombia, Costa Rica, Cyprus, Czech Republic, Denmark, Dominican Republic, El Salvador, Fiji Islands, Finland, France, Germany, Greece, Honduras, Hungary, Iceland, Indonesia, Ireland, Israel, Italy, Japan, Jordan, Korea, Latvia, Lithuania, Luxembourg, Malaysia, Malta, Mauritius, Mexico, Moldova, Morocco, Netherlands, New Zealand,  Norway, Pakistan, Panama, Paraguay, Peru, Philippines, Poland, Portugal, Romania, Russia, Saudi Arabia, Singapore, Slovenia, South Africa, Spain, Sweden, Switzerland, Thailand, Tunisia, United Kingdom, Venezuela.\\

\subsection{Correlations}
\label{app:correlations}
\added{Correlations of the explanatory variables are shown in Table \ref{tab:Correlations}. In general, correlations are quite low, with their absolute value rarely exceeding 0.35. However, political stability and regulatory quality are strongly correlated, as are GDP per capita and regulatory quality with respective correlation coefficients of 0.77 and 0.74.}
\begin{table}
	\centering
	\small
	\caption{Variable correlations.}
	\label{tab:Correlations}
	\begin{tabular}{@{}lrrrrrrrrr@{}}
		
		& \rot{GDP growth}          & \rot{Inflation CPI} & \rot{Unemployment rate} & \rot{Current account} & \rot{Government balance} & \rot{Government debt} & \rot{Political stability} & \rot{Regulatory quality} & \rot{GDP per capita}      \\ \noalign{\smallskip}\hline\noalign{\smallskip}
		GDP growth          & 1.00          & -0.19             & -0.21           & 0.09               & 0.34            & -0.24               & -0.12              & -0.10          & -0.20 \\
		Inflation CPI       &          & 1.00              & 0.17            & 0.00               & -0.20           & 0.11                & -0.05              & -0.11          & -0.03 \\
		Unemployment rate   &         &              & 1.00            & -0.20              & -0.33           & 0.20                & -0.22              & -0.23          & -0.27 \\
		Current account     &           &               &            & 1.00               & 0.22            & 0.13                & 0.05               & 0.02           & 0.13  \\
		Government balance  &          &              &            &               & 1.00            & -0.29               & 0.17               & 0.21           & 0.27  \\
		Government debt     &          &               &            &                &            & 1.00                & 0.12               & 0.15           & 0.14  \\
		Political stability &          &             &          &                &             &                 & 1.00               & 0.77           & 0.61  \\
		Regulatory quality  &          &              &           &                &             &                 &                & 1.00           & 0.74  \\
		GDP per capita      &          &            &            &               &             &                &              &           & 1.00  \\ \noalign{\smallskip}\hline
	\end{tabular}
\end{table}

\subsection{Rating transformations}
Table \ref{tab:Ratingconversion} shows the transformation of all Moody's ratings into numerical ratings.
\label{app:ratingtransformations}
\begin{table}
	\caption{Conversion of Moody's ratings into numeric ratings.}
	\label{tab:Ratingconversion}
	\begin{tabular}{@{}lr@{}}
		\hline\noalign{\smallskip}
		Moody's rating & Numeric rating                   \\
		\noalign{\smallskip}\hline\noalign{\smallskip}
		\hspace{0.5cm} Aaa & 17 \hspace{1.0cm} \\ 
		\hspace{0.5cm} Aa1 & 16 \hspace{1.0cm} \\
		\hspace{0.5cm} Aa2 & 15 \hspace{1.0cm} \\
		\hspace{0.5cm} Aa3 & 14 \hspace{1.0cm} \\
		\hspace{0.5cm} A1 & 13 \hspace{1.0cm} \\
		\hspace{0.5cm} A2 & 12 \hspace{1.0cm} \\
		\hspace{0.5cm} A3 & 11 \hspace{1.0cm} \\
		\hspace{0.5cm} Baa1 & 10 \hspace{1.0cm} \\
		\hspace{0.5cm} Baa2 & 9 \hspace{1.0cm} \\
		\hspace{0.5cm} Baa3 & 8 \hspace{1.0cm} \\
		\hspace{0.5cm} Ba1 & 7 \hspace{1.0cm} \\
		\hspace{0.5cm} Ba2 & 6 \hspace{1.0cm} \\
		\hspace{0.5cm} Ba3 & 5 \hspace{1.0cm} \\
		\hspace{0.5cm} B1 & 4 \hspace{1.0cm} \\
		\hspace{0.5cm} B2 & 3 \hspace{1.0cm} \\
		\hspace{0.5cm} B3 & 2 \hspace{1.0cm} \\ 
		\hspace{0.5cm} Caa1 & 1 \hspace{1.0cm} \\
		\hspace{0.5cm} Caa2 & 1 \hspace{1.0cm} \\
		\hspace{0.5cm} Caa3 & 1 \hspace{1.0cm} \\
		\hspace{0.5cm} Ca & 1 \hspace{1.0cm} \\
		\hspace{0.5cm} C & 1 \hspace{1.0cm} \\
		\noalign{\smallskip}\hline                      
	\end{tabular}
\end{table}

\subsection{Misclassification analysis for MLP}
\label{app:Misclassanalysis}
\added{To get further insight into the performance of the best algortihm, MLP, we perform a misclassification analysis. Results of this analysis are shown in Table \ref{tab:MLPmisclasmatrix}, where the correct predictions are on the diagonal.}

\begin{table}
	\caption{Misclassification analysis for the MLP model. Number of observations on the diagonal present the number of correct classifications by the MLP algorithm.}
	\label{tab:MLPmisclasmatrix}
	\begin{tabular}{@{}clrrrrrrrrrrrrrrrrr@{}}
		\noalign{\smallskip}\hline\noalign{\smallskip}
		\multicolumn{1}{l}{}      &                & \multicolumn{17}{c}{Credit rating MLP}                                                   \\
		\multirow{18}{*}{\rotatebox{90}{Credit rating   Moody's}} &    & 1  & 2  & 3  & 4  & 5  & 6  & 7  & 8  & 9  & 10 & 11 & 12 & 13 & 14 & 15 & 16 & 17  \\ \cline{3-19}\noalign{\smallskip}
		& \multicolumn{1}{l|}{1}  & 23 & 11 & 4  & 0  & 0  & 2  & 0  & 1  & 0  & 0  & 0  & 0  & 0  & 0  & 0  & 0  & 0   \\
		& \multicolumn{1}{l|}{2}  & 10 & 23 & 5  & 3  & 1  & 1  & 2  & 1  & 0  & 0  & 0  & 0  & 0  & 0  & 0  & 0  & 0   \\
		& \multicolumn{1}{l|}{3}  & 5  & 7  & 24 & 3  & 1  & 1  & 1  & 4  & 1  & 0  & 0  & 0  & 0  & 0  & 0  & 0  & 0   \\
		& \multicolumn{1}{l|}{4}  & 0  & 5  & 0  & 20 & 6  & 6  & 5  & 2  & 1  & 0  & 0  & 0  & 0  & 0  & 0  & 0  & 0   \\
		& \multicolumn{1}{l|}{5}  & 0  & 1  & 1  & 8  & 13 & 6  & 3  & 1  & 1  & 0  & 0  & 0  & 0  & 0  & 0  & 0  & 0   \\
		& \multicolumn{1}{l|}{6}  & 0  & 1  & 1  & 7  & 4  & 25 & 6  & 0  & 0  & 0  & 0  & 0  & 1  & 0  & 0  & 0  & 0   \\
		& \multicolumn{1}{l|}{7}  & 0  & 1  & 0  & 0  & 1  & 10 & 49 & 9  & 7  & 4  & 0  & 1  & 0  & 2  & 0  & 0  & 0   \\
		& \multicolumn{1}{l|}{8}  & 0  & 0  & 3  & 0  & 3  & 1  & 8  & 52 & 16 & 3  & 1  & 0  & 0  & 0  & 0  & 0  & 1   \\
		& \multicolumn{1}{l|}{9}  & 0  & 1  & 0  & 0  & 1  & 0  & 6  & 14 & 56 & 5  & 4  & 0  & 0  & 0  & 0  & 0  & 0   \\
		& \multicolumn{1}{l|}{10} & 1  & 0  & 0  & 0  & 0  & 1  & 3  & 6  & 6  & 39 & 7  & 1  & 1  & 1  & 0  & 0  & 0   \\
		& \multicolumn{1}{l|}{11} & 0  & 1  & 0  & 0  & 0  & 0  & 0  & 1  & 5  & 3  & 46 & 9  & 2  & 0  & 0  & 1  & 1   \\
		& \multicolumn{1}{l|}{12} & 0  & 0  & 0  & 0  & 0  & 0  & 1  & 0  & 0  & 1  & 6  & 43 & 7  & 1  & 1  & 0  & 1   \\
		& \multicolumn{1}{l|}{13} & 0  & 0  & 0  & 0  & 0  & 0  & 0  & 0  & 0  & 0  & 4  & 6  & 45 & 5  & 3  & 1  & 1   \\
		& \multicolumn{1}{l|}{14} & 0  & 0  & 0  & 0  & 0  & 0  & 2  & 0  & 1  & 0  & 1  & 1  & 4  & 26 & 0  & 1  & 2   \\
		& \multicolumn{1}{l|}{15} & 0  & 0  & 0  & 0  & 0  & 0  & 0  & 1  & 0  & 0  & 0  & 1  & 2  & 2  & 30 & 2  & 3   \\
		& \multicolumn{1}{l|}{16} & 0  & 0  & 0  & 0  & 0  & 0  & 0  & 1  & 0  & 0  & 0  & 0  & 1  & 3  & 2  & 20 & 9   \\
		& \multicolumn{1}{l|}{17} & 0  & 0  & 0  & 0  & 0  & 0  & 0  & 1  & 0  & 0  & 2  & 1  & 2  & 1  & 2  & 7  & 269 \\ \noalign{\smallskip}\hline
	\end{tabular}
\end{table}

\end{document}